\documentclass[conference]{IEEEtran}
\IEEEoverridecommandlockouts

\usepackage{cite}
\usepackage{amsmath,amssymb,amsfonts}
\usepackage{algorithmic}
\usepackage{graphicx}
\usepackage{textcomp}
\usepackage{xcolor}

\usepackage{tikz}
\usepackage{amsmath}
\usepackage{wasysym}
\usepackage{amssymb}
\usepackage{MnSymbol}
\usepackage{skak}
\usepackage{booktabs}
\usepackage{multirow}
\usepackage{makecell}
\usepackage{fontawesome5}
\usepackage{xspace,xstring}
\usepackage{lscape}

\xspaceaddexceptions{:} 
\xspaceaddexceptions{.} 
\xspaceaddexceptions{,}
\xspaceaddexceptions{+}
\xspaceaddexceptions{'}
\PassOptionsToPackage{table}{xcolor}
\usepackage{xcolor}

\usepackage{hyperref}

\usepackage{svg}
\usepackage{graphicx}
\usepackage{subcaption}
\usepackage[export]{adjustbox}
\usepackage[flushmargin]{footmisc}
\usepackage{xurl}
\usepackage{cuted}
\usepackage{stfloats} 

\usepackage{tcolorbox}
\usepackage{enumitem}
\newcommand{\roundframe}[1]{{\setlength\fboxrule{0pt}\fbox{\tcbox[colframe=black,colback=white,shrink tight,boxrule=0.5pt,extrude by=2.5pt]{\small #1}}}}
\newcommand{\rot}[1]{\multicolumn{1}{c}{\adjustbox{angle=45,lap=\width-1em}{#1}}}
\newcommand{\parvspace}{\vspace{0.20cm}}

\usepackage{amssymb}
\usepackage{xcolor}
\definecolor{capri}{rgb}{0.0, 0.75, 1.0}
\definecolor{darkgray}{rgb}{0.66, 0.66, 0.66}

\usepackage{authblk}

\newcommand{\iWhatsapp}{\includesvg[height=1em]{figures/icon_whatsapp_cropped.svg}}%
\newcommand{\iSignal}{\includesvg[height=1em]{figures/icon_signal}}%
\newcommand{\iThreema}{\includesvg[height=1em]{figures/icon_threema_quadratic.svg}}%

\newcommand{\iMessenger}{\includesvg[height=1em]{figures/icon_messenger.svg}}%
\newcommand{\iInstagram}{\includesvg[height=1em]{figures/icon_instagram.svg}}%
\newcommand{\iTiktok}{\includesvg[height=1em]{figures/icon_tiktok.svg}}%
\newcommand{\iSnapchat}{\includesvg[height=1em]{figures/icon_snapchat.svg}}%
\newcommand{\iTelegram}{\includesvg[height=1em]{figures/icon_telegram.svg}}%
\newcommand{\iViber}{\includesvg[height=1em]{figures/icon_viber.svg}}%
\newcommand{\iLine}{\includesvg[height=1em]{figures/icon_line.svg}}%

\newcommand{\eThumbup}{\includesvg[height=1em]{figures/emoji_thumb_up_u1f44d.svg}}%
\newcommand{\eHeart}{\includesvg[height=1em]{figures/emoji_heart_u2764.svg}}%

\newcommand{\serverAck}{\includegraphics[height=1em]{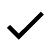}}%
\newcommand{\deviceAck}{\includegraphics[height=1em]{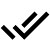}}%
\newcommand{\readAck}{\includegraphics[height=1em]{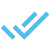}}%

\usepackage[colorinlistoftodos,prependcaption,disable]{todonotes} %

\newcommand{\aju}[1]{\todo[linecolor=blue,backgroundcolor=blue!25,bordercolor=blue,inline]{\textbf{Aljosha:} #1}\noindent}
\newcommand{\gge}[1]{\todo[linecolor=yellow,backgroundcolor=yellow!25,bordercolor=yellow,inline]{\textbf{Gabriel:} #1}\noindent}

\newcommand{\creepy}{creepy companion\xspace}
\newcommand{\spooky}{spooky stranger\xspace}
\newcommand{\creepys}{creepy companions\xspace}
\newcommand{\spookys}{spooky strangers\xspace}
\newcommand{\Creepy}{Creepy companion\xspace}
\newcommand{\Spooky}{Spooky stranger\xspace}

\newcommand{\Spookys}{Spooky strangers\xspace}

\newcommand{\CREEPYS}{Creepy Companions\xspace}
\newcommand{\SPOOKYS}{Spooky Strangers\xspace}

\newcommand{\seconds}{\nobreak\hspace{.16667em plus .08333em}\nobreak s\xspace}
\newcommand{\mseconds}{\nobreak\hspace{.16667em plus .08333em}\nobreak ms\xspace}

\def\BibTeX{{\rm B\kern-.05em{\sc i\kern-.025em b}\kern-.08em
    T\kern-.1667em\lower.7ex\hbox{E}\kern-.125emX}}
\begin{document}

\title{Careless Whisper:\\Exploiting Silent Delivery Receipts to Monitor Users on Mobile Instant Messengers
}

\makeatletter
\renewcommand\AB@affilsepx{, \protect\Affilfont}
\setlength{\affilsep}{0.5em}   %
\makeatother

\author[1,2]{Gabriel K. Gegenhuber}
\author[1]{Maximilian Günther}
\author[1]{Markus Maier}
\author[1]{\authorcr Aljosha Judmayer}
\author[1]{Florian Holzbauer}
\author[3]{Philipp É. Frenzel}
\author[1]{Johanna Ullrich}

\affil[1]{University of Vienna, Faculty of Computer Science}
\affil[2]{UniVie Doctoral School Computer Science}
\affil[3]{SBA~Research}

\pagestyle{plain}

\maketitle

\begin{abstract}

With over 3 billion users globally, mobile instant messaging apps have become indispensable for both personal and professional communication.
Besides plain messaging, many services implement additional features such as delivery and read receipts informing a user when a message has successfully reached its target.
This paper highlights that delivery receipts can pose significant privacy risks to users. We use specifically crafted messages that trigger silent delivery receipts allowing any user to be pinged without their knowledge or consent.
By using this technique at high frequency, we demonstrate how an attacker could extract private information such as following a user across different companion devices, inferring their daily schedule, or deducing current activities.
Moreover, we can infer the number of currently active user sessions (i.e., main and companion devices) and their operating system, as well as launch resource exhaustion attacks, such as draining a user's battery or data allowance, all without generating any notification on the target side.
Due to the widespread adoption of vulnerable messengers (\emph{WhatsApp} and \emph{Signal}) and the fact that any user can be targeted simply by knowing their phone number, we argue for a design change to address this issue.

\aju{In addition to read receipts, delivery receipts are a design inherent and confirm the correct processing of the message by the receiving device.}
\aju{By using this technique at high frequency, we demonstrate how an attacker can monitor a user's device state throughout the day, enabling inferences about their behavior and daily routine. 
This includes detecting transitions between Wi-Fi and cellular networks, or identifying the use of different end devices—such as switching between a mobile phone and a desktop, as well as their respective operation systems. For certain mobile devices, our fine-grained timing measurements even allow an attacker to determine whether the screen is currently on or off. These findings reveal that delivery receipts can pose a significantly greater privacy risk than previously recognized.
Moreover, we describe resource exhaustion attacks via this probing channel, such as draining a user's battery or data allowance, all without generating any notification on the target side.
}

\end{abstract}

\title{Careless Whisper: Exploiting Stealthy End-to-End Leakage in Mobile Instant Messengers}

\section{Introduction}
Instant messengers serve a vast global audience with WhatsApp alone reaching over 3 billion users~\cite{noauthor_whatsapp_about, gegenhuber_2025_heythere} and handling billions of messages daily. 
Besides being very common in general, instant messaging services are also used by high-profile government officials for sensitive conversations~\cite{atlantic_2025_houthi, spiegel_private_2025}, which adds an entirely different dimension to privacy issues within these services.
In this paper, we present a novel privacy and availability attack vector on instant messaging systems, leveraging the (ab)use of \emph{delivery receipts}.

There are two basic ways used by instant messengers to inform senders about message delivery, namely \textit{delivery receipts} (consisting of \textit{server ack \& device ack)} and \textit{read receipts}. 
The first acknowledges a message's receipt at the server or the destination device, the latter their view by the destination device's user.
Read receipts have been misused to spy on conversation partners~\cite{freed_stalker_2018} and nowadays messenger applications allow to disable them in their privacy settings. 
Delivery receipts, however, cannot be deactivated due to design choices of the underlying protocol.
Previous work has triggered delivery receipts through sending regular text messages in ongoing conversations and thereby showed that, based on the measured round-trip times (RTTs), country-level geolocation of a user's device is feasible ~\cite{schnitzler_hope_2023}.
Regular messages, however, trigger notifications for the target user, potentially alerting them to the ongoing attack, particularly when the probing messages are sent frequently.

Using techniques described in this paper, an adversary can craft stealthy messages that enable probing a target at high frequency (up to sub-second granularity) while not causing any notification at the target side and also in the absence of an ongoing conversation. 
With such an increased sampling rate, we systematically show that delivery receipts can leak user information beyond the country level.

\begin{figure}[t]
  \centering
  \includegraphics[width=\linewidth]{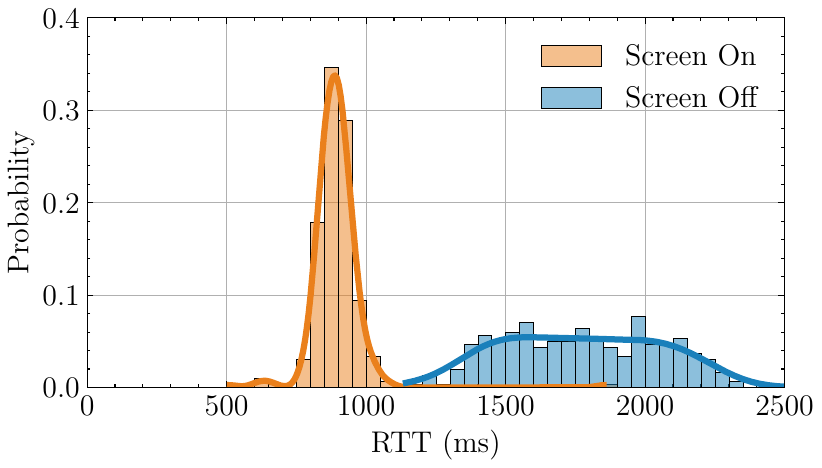}
  \caption{Round-trip times (RTT) of delivery receipts, which are $\le1$ second for \textit{Screen On} states and $>1$ second and above for \textit{Screen Off} states measured on an iPhone using WhatsApp with a sampling rate of 1 Hz.}
  \label{fig:density-graph-iphone}
\end{figure}

For example, we show that the on/off state of a mobile phone's screen manifests in the delivery receipts' timing,
see Figure~\ref{fig:density-graph-iphone},
and, among others, allows to track the victim's screen time.

Moreover, we demonstrate that a user's activity can be followed across multiple devices (i.e., smartphone and companion sessions), creating further (and more severe) monitoring and tracking possibilities.
In addition to utilizing delivery receipt RTT as a timing side channel, we demonstrate that implementation inconsistencies across different target architectures also leak information about the operating systems and application clients in use.

Using the same techniques, also resource exhaustion attacks, such as draining the battery or data quota, can be performed.
Similar to \cite{schnitzler_hope_2023} we focus our analysis on three instant messenger platforms, i.e., WhatsApp and the more security-orientied alternatives Signal and Threema.

In summary, we make the following contributions:

\parvspace
\noindent\textbf{Stealthy Delivery Receipts.} While previous work required an ongoing conversation and consequently notified the victim about every probing message, we show that delivery receipts are also issued for other message types (e.g., reactions) and furthermore can be triggered in a silent way preventing a notification of the victim. This allows \textbf{constant} and \textbf{high-frequency probing} (i.e., sub-second interval) of target devices without the risk of getting noticed and blocked by the victim.

\parvspace
\noindent\textbf{Multi-Device Amplification.} We demonstrate that in multi-device setups -- in which victims use web or desktop clients for messaging in addition to their mobile phones -- each device can be independently probed enabling comprehensive and precise observation of user behavior across devices throughout the day.

\parvspace
\noindent\textbf{Arbitrary Targets.} For WhatsApp and Signal, we show that it is not even required to have any kind of association with the victim, e.g., being in their contact list or having an existing conversation. Thereby, anyone having these messenger applications installed on their mobile phone can be selected as a victim just by knowing their phone number and monitored using the techniques described in this paper. 
With billions of users, the  number of potential victims is not only huge but also includes potential high profile targets such as government officials~\cite{atlantic_2025_houthi, spiegel_private_2025, whittaker_encryption_2024, cerulus_eu_2020}. %

\parvspace
\noindent\textbf{Activity Leakage.} We show that delivery receipt timings are influenced by the phone's activity and sleep states. This enables us to determine whether the remote device is actively used or in standby by distinguishing between screen on/off states and in certain scenarios even if the messaging app is currently in foreground. Furthermore, we demonstrate that complex routines, schedules, and activities can influence delivery receipt timings across a user's different devices, as shown by an open-world measurement case conducted under real-world conditions.

\parvspace
\noindent\textbf{Resource Exhaustion Attacks.} We show that our findings can not only be used for the disclosure of private information, but also in an offensive way exhausting a victim's resources like battery or data quota -- similarly, this type of attack does not alert the victim through notifications.

\parvspace
\noindent\textbf{Countermeasures.} We discuss how these attacks can be mitigated, including both client and service-side mitigation strategies.

\section{Background}

In prevalent end-to-end encrypted (E2EE) messaging protocols used in WhatsApp and Signal, the role of the server is essentially reduced to forwarding encrypted messages to their recipients, and large parts of the protocol logic are shifted to the client side. 
This also includes the handling of re-encryptions and re-submission in case of decryption failures at the receiver, e.g., if the session state and associated keys on a recipient device have been deleted, or rolled back.
Thus, in addition to serving as a convenience feature for users, clients also depend on information about the successful delivery and decryption of messages from a technical standpoint.
These acknowledgments are commonly referred to as \textit{delivery receipts}. 
Delivery receipts indicate the successful decryption of a message and thus allow 
the sender to mark the transmitted message and the associated ephemeral keying material for deletion.
This is necessary to uphold the security property of forward secrecy as promised by these messaging services. 
As we show in this paper, this design decision with the according shift of responsibilities to clients in combination with the desired responsiveness and convenience of low latency interactions can have a significant impact on the privacy as well as the security of users. 

\begin{figure}
    \centering
    \includegraphics[width=0.9\linewidth]{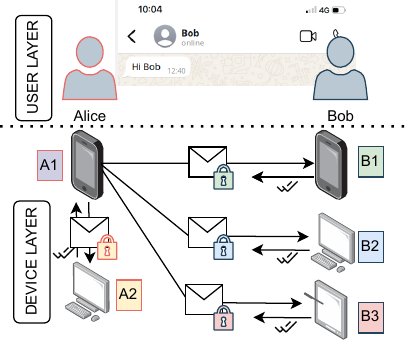} 
    \caption{Simplified depiction of client-fanout for Multi-Device-Support: Alice's message is sent to all of Bob's devices as well as her desktop device. Each message copy is individually encrypted. The recipient devices inform Alice's device of the successful decryption via delivery receipts.}
    \label{fig:multiplicationmultidevice}
\end{figure}

\parvspace
\noindent\textbf{Message Delivery.}
The \textit{server ack} delivery receipt, usually represented by a single checkmark (\serverAck) on client devices, indicates that the message was queued for further transmission at the message server. Due to E2EE, the message server does not see the message content and thus cannot perform message validation checks in this step. %
The \textit{device ack} delivery receipt, often represented by two checkmarks (\deviceAck) on client devices, highlights that the recipient has received and successfully decrypted the message. 
Therefore, the delay between sending the message and receiving the delivery receipts indicates the time it took for sending the message first to the message server, then from the message server to the target device (if it is currently online) and back again. 
For details regarding the message server infrastructure, the resulting RTTs and the inferred limits for geolocation see %
Appendix \hyperref[sec:infra]{A1}.
The \emph{read receipt} (\readAck) is a message type returned by the target device in case the target user accesses the message. 
However, \mbox{WhatsApp}, Signal and Threema allow to disable this message type in the device settings.

\parvspace
\noindent\textbf{Multi Device Setups.}
There exist \emph{leader-} and \emph{client-fanout} based approaches to support multi device (i.e., companion) setups.
In contrast to leader-based approaches, where one device acts as a primary device accountable for redistributing messages to other devices, current client-fanout implementations trigger delivery receipts by \emph{all} user devices connected to the account. 
This client-fanout is now implemented in WhatsApp and Signal. 
Threema is currently in the process to enable multi-device setups~\cite{threema_ios_multidevice}.
In client-fanout setups all devices of the user maintain their own key pair, cf. Figure~\ref{fig:multiplicationmultidevice}.  
For messaging, the sender creates an individual E2EE channel with \emph{each} device of the receiver as if messaging multiple recipients. 
For remaining consistency among all devices of the sender, the sending device also forwards the message (and other information) to other devices of the same user~\cite{marlinspike_sesame_2017}.
This approach avoids the single point of failure inherent in the leader-based method. 
However, assigning multiple keys to an account reveals the number of devices under the control of a user, as other accounts necessarily have to retrieve these keys from a central inventory. 
Since each device has its own unique key, the recipient can also infer from which of the sender’s devices the message originated~\cite{beery_whatsapp_2024}.

\parvspace
\noindent\textbf{Mitigation Status.}
Although previous work \cite{schnitzler_hope_2023} has shown that delivery receipts introduce a timing-based side channel that coarsely leaks a user's location or the used access technology (i.e., cellular, Wi-Fi), there is still no way of turning them off at most popular messengers.
Moreover, other proposed mitigation techniques like delaying the delivery receipt by a random delay of a few seconds were not adopted.

\section{Threat Model \& Measurement Setup}
\label{sec:threat-and-attacker-model}

\begin{table}[t]
    \centering
    \newcommand{\trow}[6]{#1 & #2 & #3 & #5 & #4 & #6 \\}
    \newcommand{\head}[1]{\rot{\textbf{#1}}}%
    \newcommand\tFD{$^{\dag}$}
    \newcommand\tFDD{$^{\ddag}$}
    \newcommand\tFa{$^{\mathrm{a}}$}
    \newcommand\tFb{$^{\mathrm{b}}$}
    \newcommand\tFc{$^{\mathrm{c}}$}
    \begin{tabular}{@{}lrcccc@{}}
        \trow{\head{Application}} {\head{Installations}} {\head{E2EE Messaging}} {\head{Delivery Receipts}} {\head{Multi Device E2EE}}  {\head{Open Source}} \midrule
        \trow{\iWhatsapp~WhatsApp}     {9.64B}  {\CIRCLE}      {\CIRCLE}  {\CIRCLE}     {\Circle}
        \trow{\iMessenger~Facebook Messenger}    {5.89B}  {\CIRCLE}      {\CIRCLE}  {\CIRCLE}     {\Circle}
        \trow{\iInstagram~Instagram}    {5.48B}  {\CIRCLE}      {\Circle}  {\CIRCLE}     {\Circle}
        \trow{\iTiktok~TikTok}       {3.94B}  {\Circle}      {\Circle}  {\Circle}     {\Circle}
        \trow{\iSnapchat~Snapchat\tFa} {1.94B}  {\Circle}      {\CIRCLE}  {\Circle}     {\Circle}
        \trow{\iTelegram~Telegram\tFb} {1.64B}  {\LEFTcircle}  {\Circle}  {\Circle}     {\CIRCLE}
        \trow{\iViber~Viber}        {1.18B}  {\CIRCLE}      {\CIRCLE}  {\CIRCLE}     {\Circle}
        \trow{\iLine~Line}         {1B}     {\CIRCLE}      {\Circle}  {\Circle}     {\Circle}
        \hline
        \trow{\iSignal~Signal}       {136M}   {\CIRCLE}      {\CIRCLE}  {\CIRCLE}     {\CIRCLE}
        \trow{\iThreema~Threema\tFc}  {5M}     {\CIRCLE}      {\CIRCLE}  {\LEFTcircle} {\CIRCLE}
        \bottomrule  
    \end{tabular}
    \\
    \vspace{1mm}
    \tFa~E2EE for snaps, not for private messages. \\
    \tFb~Non-default E2EE.\\ %
    \tFc~Multi-device support for iOS only. \\ %
    \caption{Overview over popular ($>$1B installations) and security-oriented instant messaging services (Sources: androidrank.org, manufacturer information).}
    \label{tab:messenger-overview}
\end{table}

We assume an adversary that aims to impinge a victim's privacy by exploiting instant messenger services' delivery receipts.
More specifically, we differentiate between i) three distinct attacker goals, ii) two attacker types differing in their relation to the victim and iii) limit our analysis to three popular messaging services -- WhatsApp and its more security-oriented alternatives Signal and Threema.

\subsection{Attack Goals}
From an adversary's perspective, the three goals are:
\begin{itemize}\parskip=3pt plus 3pt\itemsep=0pt
    \item[\roundframe{G1}] Fingerprinting the number and types of a victim's devices by observing received messages and delivery receipts, thereby inferring each device's online status and enabling tracking across multiple devices, locations, and behavioral routines.
    \item[\roundframe{G2}] Monitoring a user's behavior (e.g., screen on/off, or messenger app currently in foreground) by covertly probing the device for longer periods and analyzing relative RTT differences of delivery receipts.
    \item[\roundframe{G3}] Launching offensive resource exhaustion attacks that increase a user's traffic, draining their battery or data allowance, or performing denial-of-service attacks to degrade the usability of bandwidth-intensive applications such as video calls on secure messaging platforms.
\end{itemize}

The adversary aims to carry out the attack as stealthily as possible, ensuring that the victim remains unaware not only of the source and cause, but also of the fact that an attack is occurring at all.

\subsection{Attacker Types}
Depending on the relationship between the adversary and the victim, we differentiate between two attacker types, namely \emph{\creepy} and \emph{\spooky}.
\begin{itemize}
    \item \textbf{\emph{\Creepy}}:
    The attacker and the victim have an active chat session containing one or more messages in the messaging app.
    Such adversaries, e. g., a jealous (ex-)partner or a nosy employer,  typically also have real-life relationship with the victim.
    \item \textbf{\emph{\Spooky}}:
    There is no prior relationship -- like  a previous preexisting contact or conversation -- between the attacker and the victim in the instant messaging app.
    The adversary only knows the latter's phone number.
    This way, any customer of a messaging service could be attacked just by knowing their phone number.
    This could be used to spy on public figures, e. g., celebrities or politicians, or 
    to gather intelligence about a company's CEO for industrial espionage.
    \emph{\Spookys} can simply use a so-called burner phone with a prepaid SIM card to entirely hide their identity against the messaging service.

\end{itemize}

\subsection{Messenger Selection}
Table~\ref{tab:messenger-overview} shows the most popular instant messenger applications including also the social networks
Instagram, TikTok, and Snapchat as they also support direct messaging. 
At the table's bottom, we listed the less prevalent but more security-oriented messengers Signal and Threema.
For our analysis,
we require messengers to support multi-device E2EE and delivery receipts limiting the potential target applications to five, namely WhatsApp, Facebook Messenger, Viber, Signal, and Threema. 
Beyond, we need API-level access.
While this is clearly available for open-source messengers,
the communication protocol and API endpoints need to be reverse engineered for proprietary messengers.
In our research,
we found web gateways enabling messaging as a non regular user, e.g., the Viber REST API\footnote{\href{https://developers.viber.com/docs/api/rest-bot-api/}{\texttt{developers.viber.com/docs/api/rest-bot-api}}},
and projects emulating a webbrowser providing automation capabilities but no in-depth access to message delivery states.
Finally, we only considered projects that implement the full feature set of a regular client, support E2EE key management and expose low-level API access, see Table~\ref{tab:overview-api-access} for an overview.
This left us with three messengers for further analysis, \emph{WhatsApp}, \emph{Signal} and \emph{Threema}.
This choice of messengers is consistent with~\cite{schnitzler_hope_2023} and allows comparison with previous work.

\subsection{Measurement Setup}
We conducted our measurements in two steps:
First, we looked at message types triggering delivery receipts (on the iOS and Android apps) without looking at specific device models and evaluate them based on our threat model, see Section~\ref{sec:side-channel-discovery} for details.
Based on this assessment, we craft attacks on the individual messengers and test the privacy leaks for specific manufacturers and models in a second step, see Section~\ref{sec:attacks} details. The full list of our testing devices including their chipsets and software versions is shown in Table~\ref{tab:testing-devices} in the Appendix.

\begin{table}[]
    \centering
    \newcommand\tO{$^{\mathrm{\symking}}$}
    \newcommand\tC{$^{\downpitchfork}$}
    \begin{tabular}{@{}ll@{}}
    \toprule
    \textbf{App} & \textbf{Used Open Source Projects\footnotemark}     \\ \midrule
    \iWhatsapp~WhatsApp   & whatsmeow\tC (web), Cobalt\tC (mobile) \\
    \iSignal~Signal      & signal-cli\tC (uses libsignal\tO)      \\
    \iThreema~Threema    & threema-android\tO                     \\ \bottomrule
    \end{tabular}
    \\
    \tO~Offical project. \hspace{1ex}
    \tC~Community project (reverse-engineered).
    \caption{Both official- and community-driven open source projects were leveraged to get API level application access.}
    \label{tab:overview-api-access}
\end{table}
\footnotetext{
whatsmeow: %
\href{https://github.com/tulir/whatsmeow}{\texttt{github.com/tulir/whatsmeow}}\\
Cobalt: %
\href{https://github.com/Auties00/Cobalt}{\texttt{github.com/Auties00/Cobalt}}\\
signal-cli: %
\href{https://github.com/AsamK/signal-cli}{\texttt{github.com/AsamK/signal-cli}}\\
libsignal: %
\href{https://github.com/signalapp/libsignal-service-java}{\texttt{github.com/signalapp/libsignal-service-java}}\\
threema-android: %
\href{https://github.com/threema-ch/threema-android}{\texttt{github.com/threema-ch/threema-android}}
}

\section{Side Channel Vectors}
\label{sec:side-channel-discovery}

To explore the probing capabilities and corresponding limitations for exploitation, we tested for side channels based on delivery receipts.
More specifically, we analyzed
i)~which actions cause delivery receipts,
ii)~whether delivery receipts are also issued for messages coming from a \emph{\spooky}, %
iii)~how delivery receipts are issued within multi-device setups.

\subsection{Delivery Receipt Sources}
We systematically tested which actions trigger delivery receipts by using our custom clients to send messages to both Android and iOS phones.
Moreover, we examined which actions notify the user (e.g., trigger a push notification or mark a conversation as unread) and which actions remain covert. Due to their stealth, the latter bear the potential of continuous monitoring of target users without them being notified by the messaging app.

More specifically, we systematically test and explore delivery receipts caused by the following actions:
\begin{itemize}
    \item \textbf{Send message}: sending a normal (text) message to the target.
    \item \textbf{Edit message}: changing the content of a previously sent message.
    \item \textbf{React to message}: sending a message reaction (e.g., a~\eThumbup \xspace or~\eHeart~emoji) to an existing message.
    \item \textbf{Delete message}: revoking a previously sent message for all chat participants (\textit{``delete for everyone''}).
\end{itemize}

\begin{table}[]
\centering
\newcommand\tWhatsapp{\tikz[baseline=(O.base)]{\node(O) [baseline,minimum width=6mm,inner sep = 0,align=left] {\iWhatsapp};}}
\newcommand\tSignal{\tikz[baseline=(O.base)]{\node(O) [baseline,minimum width=6mm,inner sep = 0,align=left] {\iSignal};}}
\newcommand\tThreema{\tikz[baseline=(O.base)]{\node(O) [baseline,minimum width=6mm,inner sep = 0,align=left] {\iThreema};}}

\begin{tabular}{l ccc ccc}
\toprule
\textbf{Action} & \multicolumn{3}{c}{\textbf{Delivery Receipt}} & \multicolumn{3}{c}{\textbf{Push Notification}} \\
                & \tWhatsapp      & \tSignal     & \tThreema     & \tWhatsapp      & \tSignal        & \tThreema  \\
                \midrule
Message         & \CIRCLE         & \CIRCLE      & \CIRCLE       & \CIRCLE         & \CIRCLE         & \CIRCLE    \\
Reaction        & \CIRCLE         & \CIRCLE      & \Circle       & \LEFTcircle     & \LEFTcircle     & \Circle    \\
Edit            & \CIRCLE         & \CIRCLE      & \Circle       & \faApple        & \Circle         & \Circle    \\
Delete          & \CIRCLE         & \CIRCLE      & \Circle       & \Circle         & \Circle         & \Circle    \\ \bottomrule
\end{tabular}
\\
\faApple~Edits cause (silent) notifications for iOS users only (no notifications are shown on Android).\\
\caption{Different actions notify the sender via delivery receipt and the receiver via push notification.
On WhatsApp and Signal, reactions only cause push notifications for messages originated by the receiver but not those by other users (hence marked with \LEFTcircle).
For edits and deletions, WhatsApp and Signal employ restrictions (i.e., time window, recurrence).
}
\label{tab:receipts-and-stealthiness}
\end{table}
Table~\ref{tab:receipts-and-stealthiness} shows the results.
While WhatsApp and Signal also send delivery receipts for reactions, edits, and message deletions, Threema restricts the delivery receipts to regular messages.
Editing or deleting a message usually does not trigger a notification on the target's phone and could thus be used for tracking purposes.
However, both WhatsApp and Signal impose restrictions on these actions. WhatsApp permits message deletion for up to two days
and allows unlimited edits within 15 minutes.
In Signal, the time frame for deleting and editing messages is 24 hours with an upper limit of 10 edits per message.

\subsection{Stealthy Probing (\textit{\CREEPYS})}
\label{subsec:stealthy-creepy}
Threema only allows reacting to someone else's message but users on WhatsApp and Signal can also send reactions to their own messages.
A user is only notified when somebody reacts to a message originally sent by them.
Self-reactions do not provoke a notification for other chat participants but nevertheless trigger a delivery receipt. 
Therefore, self-reactions provide an inconspicuous way of probing a target to receive delivery receipts.
To make things worse, there are no time or quantity restrictions on message reactions and users can change or remove their reaction at a later point in time.
Thus, an attacker could simply react to an old message they sent themselves to stay under the radar.
Finally, removing a reaction (i.e., sending an empty string as a message reaction) is entirely invisible to the targeted user providing an ideal vector for consistent monitoring.

The just described side channels do not necessarily require full API access.
An attacker could use an official client, or manipulate the client's state (e.g., by using developer tools via the web app) to trigger inconspicuous reactions and observe the (encrypted) traffic to derive delivery report timings\footnote{This technique is used in~\cite{schnitzler_hope_2023}. In contrast to this work, the authors exploit delivery receipts of regular messages triggering notifications at the victim device. This implies that their probing is not stealthy and cannot be applied in a continous and high-frequent way as we do.}.
However, having API-level access facilitates the probing and shows that invalid messages are generously confirmed via delivery receipts.
For example, a message deletion packet can be sent multiple times to harvest continuous delivery receipts (with only the first message deletion actually being effective).
Although message deletion or edits that were sent after the official time window were not considered (i.e., executed) by the receiving client\footnote{Officially announced times differ from the actually enforced ones, e.g., deleting on WhatsApp is possible for 60 hours instead of 48 hours and editing for 20 minutes instead of 15 minutes. For Signal, the observed time window is 48 hours instead of 24 hours.}
these messages also generated delivery receipts, providing another stealthy side channel that could be used for tracking purposes.

Summing up, we show that an attacker (more specifically a \emph{\creepy}) can use the \mbox{\textit{remove~reaction}} action, or reactions to their own messages to stealthily monitor any target that has an existing conversation with them. 
The only requirement is a conversation with at least one message for the target that is then used by the attacker for reactions.

\subsection{Stealthy Probing (\textit{\SPOOKYS})}
\label{subsec:stealthy-spooky}

Our tests showed that there is little validation done by the receiving client and that message reactions referring to non-existing messages also trigger delivery receipts.
This removes the prerequisite of having an existing conversation containing a message that a reaction refers to.
Therefore, this could also be exploited by \emph{\spookys}.

To explore this scenario we purchased a new prepaid SIM, plugged it into a burner phone, and again used our custom clients to probe various target phones that do not have any previous relation (e.g., contact, conversation, group chat) with the attacker's phone number.

Our results showed that both WhatsApp and Signal allow arbitrary targets to be stealthily monitored by a \emph{\spooky} via reactions referencing non-existing messages.
Due to missing self-reactions, we did not identify a covert way of probing arbitrary targets on Threema as a \emph{\spooky}.

\subsection{Multi-Device Probing}
\label{subsec:multi-device-amplification}
Besides observing which actions can be used to probe a target user via delivery receipts, we also analyzed how delivery receipts are handled when the target uses multiple devices.
For both, \mbox{WhatsApp} and Signal, all devices (Android, iOS, Mac, Windows, Linux, and Web) issue independent (i.e., duplicated) delivery receipts for all tested message types.
If the device is online, delivery receipts are issued right away;
if not, they are sent as soon as the connectivity of the device is regained.
This amplification of delivery receipts further increases an attacker's tracking and fingerprinting possibilities (as shown in Section~\ref{subsec:monitoring-device-online}) as they are capable of inferring when a device comes online from the receipt of pending delivery receipts.
On Threema, we checked for multi-device receipts by sending normal text messages.
Threema appears to synchronize issued delivery receipts among all devices of a user causing only a single delivery receipt per message.

\subsection{Summary of Probing Capabilities}

We summarize the identified side channels that can be used for covert probing of user devices in Table~\ref{tab:stealthy-probing}. 
Threema only responds with a single delivery receipt even in settings with multiple devices per user.
Moreover, Threema does not allow a \emph{\spooky} or \emph{\creepy} to covertly probe a user's device without triggering notifications for the victim. 
The only way that remains to trigger delivery receipts by a previously unknown user is by sending a normal text message and starting a new conversation, 
but this is obviously not stealthy.
Summarizing, Threema handles delivery  receipts in a restrictive way impeding stealth probing. Consequently, we focus on the the messaging  services WhatsApp and Signal in the remainder of this paper.

\begin{table}[]
\centering
\newcommand\tWhatsapp{\tikz[baseline=(O.base)]{\node(O) [baseline,minimum width=6mm,inner sep = 0,align=left] {\iWhatsapp};}}
\newcommand\tSignal{\tikz[baseline=(O.base)]{\node(O) [baseline,minimum width=6mm,inner sep = 0,align=left] {\iSignal};}}
\newcommand\tThreema{\tikz[baseline=(O.base)]{\node(O) [baseline,minimum width=6mm,inner sep = 0,align=left] {\iThreema};}}

\begin{tabular}{l c c c}
\toprule
\thead{\textbf{Messenger used for}\\\textbf{covert probing}} & \thead{\textbf{Spooky} \\ \textbf{stranger}} & \thead{\textbf{Creepy} \\ \textbf{companion}} & \thead{\textbf{Each} \\ \textbf{Device}}  \\
\midrule
\tWhatsapp~WhatsApp   &    yes      & yes         & yes    \\
\tSignal~Signal       &    yes      & yes         & yes    \\
\tThreema~Threema     &    no       & no          & no    \\ \bottomrule
\end{tabular}
\caption{Ability to covertly probe a target, i.e, without triggering a notification, using delivery receipts.
The last column specifies if this is possible for each of a user's devices individually.
}
\label{tab:stealthy-probing}
\end{table}

\section{Attacks \& Exploitation}
\label{sec:attacks}
In this section, we show the manifold potential for abuse and exploitation of the discovered delivery receipt-based side channels for the two vulnerable applications WhatsApp and Signal.
More specifically, we use stealthy message reactions to extract privacy-sensitive information from and for offensive resource exhaustion attacks against the victim.
Most of the presented exploits require the attacker to continuously send stealthy message reactions towards the victim. %

\subsection{Tracking Users Across Devices}
\label{subsec:tracking-users-across-devices}
Concurrent work~\cite{beery_whatsapp_2024} shows that WhatsApp's key directory leaks the device setup information of users and that the used sender device can be extracted from received messages.
We validated these findings and confirmed that this is also the case for Signal.
For WhatsApp and Signal, the main device has the lowest device index (0 and 1 respectively) enabling differentiation between main- and companion devices.
Monitoring a user's device directory and consequently observing a user's number of devices can be executed by \emph{\spookys}.

\begin{figure}
    \centering
    \includegraphics[width=\linewidth]{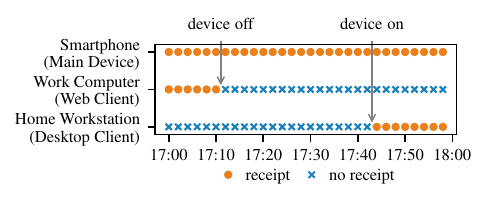}
    \caption{A device's online status can be consistently and stealthily monitored with second-based granularity, possibly leaking the user's location and daily routines.}
    \label{fig:companion-online-status-monitoring}
\end{figure}

\subsection{Monitoring Device Online Status}
\label{subsec:monitoring-device-online}
Besides monitoring a target's key directory on the server, we are able to actively send packets that remain hidden to the victim.
As all devices answer individually with a delivery receipt,
continuous probing enables independent monitoring of each device's online status.
In this use case, the attacker does not evaluate the RTT of the delivery receipts but simply their time of receipt to trace the online state of a device.

Main devices, i.e., mobile phones, are expected to be online most of the time, either via Wi-Fi or a cellular connection.
If the main device ceases to respond, 
an attacker might deduce a brief connection disruption or that the phone has been switched to airplane mode.
By monitoring this status over a longer period of time, 
an attacker might then be able to extrapolate the victim's behavioral patterns.
For example, absent delivery receipts could indicate the victim being on a flight, or at a location with no coverage, e.g., in the metro, elevator, or basement, or simply using the phone's airplane mode to mute all messages during the night revealing their sleep schedule.

Companion devices (desktop or web clients) return delivery receipts upon missed messages as soon as they come online and the adversary will automatically be notified.
This behavior might be abused to track the victim across devices and potentially expose their location, e.g., in case the companion device is a desktop computer in the office or at home (cf. Figure~\ref{fig:companion-online-status-monitoring}).
While web sessions need to be initiated by opening the corresponding website in the browser, desktop clients are often automatically started as a system service allowing an attacker to precisely monitor the online status of a companion device.
As presented in the previous section, WhatsApp and Signal allow stealthy and independent pinging of all existing devices allowing both \emph{\creepys} and \emph{\spookys} to exploit this attack.

\begin{figure}[t]
  \centering
  \includegraphics[width=.8\linewidth,trim={2.5mm 2.5mm 2.5mm 2.5mm},clip]{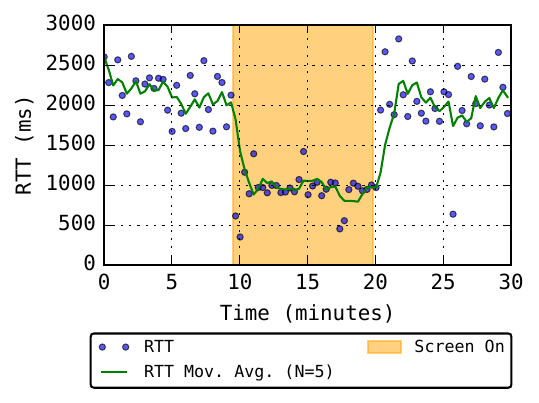}
  \caption{WhatsApp Screen On/Off: Measured with low frequency (1 ping per 20\seconds), RTTs enable to differentiate between inactive and active screen states.}
  \label{fig:whatsapp-ios-screen-on-off}
\end{figure}

\subsection{Fingerprinting User Behavior}
\label{subsec:fingerprinting-user-behaviour}
Besides using delivery receipts to classify a device's (binary) online status, relative differences in the observed RTTs can be used to derive the activity of the target device.
We found that the operating system, the smartphone's model, its underlying chipset, and the current environment (e.g., screen- and target application status or Wi-Fi vs. LTE) heavily influence the occurring RTTs for a device.

\parvspace
\noindent\textbf{Ping Frequency.}
We use different probing frequencies to measure characteristic RTTs on testing devices.
On WhatsApp, we did not experience any rate limiting or server-side queuing and could also send high-frequency ping messages, e.g., one reaction every 50\mseconds, without any restrictions.
On Signal, short bursts were also permitted but sending multiple messages per second continuously over an extended period caused them to queue. Thus, we refrained from sending more than one ping per second to circumvent rate limiting.

\subsubsection{Case Study: iPhone}
\label{subsec:case-study-iphone}
To investigate the feasibility of extracting detailed statistics about the ongoing user activity (e.g., screen time and app activity state), we systematically measured Android and iOS phones in different environments.
Due to its dominant market share (currently, more than 50\% in the US~\cite{statcounter_mobile_2024}), we select the iPhone to showcase our results.
In our tests, we compared two different iPhone models (iPhone 13 Pro, iPhone 11), both showing the same characteristic patterns for specific activities (screen on/off, application in foreground).
For our systematic measurements, we fixed the ping rate to one packet every 2, or 20 seconds.

Complete graphs of our systematic measurements comparing different attacker types (\emph{\spooky} vs. \emph{\creepy}), access technologies (Wi-Fi vs. LTE), Applications (WhatsApp vs. Signal), and iPhone models (iPhone 13 Pro, iPhone 11) can be found in Appendix~\hyperref[sec:appendix-iphone]{A5} and proof the feasibility of our monitoring across all tested environments.
We also uploaded a video\footnote{\url{https://drive.proton.me/urls/DHACRYX250\#CLYkdE3Rb7Ho}} demonstrating a single measurement case (\emph{\creepy}, 20\seconds interval, Wi-Fi on the iPhone 11, video speedup 30x).
Initially, we conducted our measurements manually.
However, we later automated the process using an ESP32, which emulates a Bluetooth keyboard and executes the keyboard inputs to switch between the required activity states.

For clarity and ease of demonstration, we isolate particular patterns and present graphs focusing on these findings under fixed conditions (i.e., a \emph{\creepy} using WhatsApp to track a target connected via Wi-Fi). However, complementary measurements (Appendix~\hyperref[sec:appendix-iphone]{A5}) showed that these patterns generalize and can similarly be observed under varying conditions.

\begin{figure}[t]
  \centering
  \includegraphics[width=.82\linewidth,trim={2.5mm 2.5mm 2.5mm 2.5mm},clip]{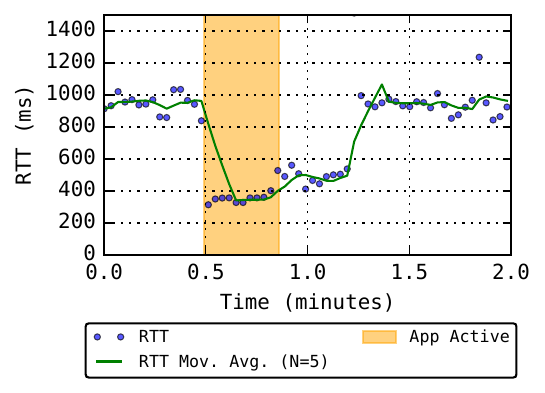}
  \caption{WhatsApp Use:  RTTs are~350\mseconds if the application is active (foreground). If minimized, RTTs become 500\mseconds for 30\seconds before eventually returning to 1\seconds \textit{screen on} as typical for long-term app standby.}
  \label{fig:whatsapp-ios-app-active-inactive}  
\end{figure}

\subsubsection*{Showcase I: Deriving Smartphone Screen Time}
While knowing (one own's) screen time is convenient for digital wellbeing and parental control, it is also interesting for external entities.
For example, a nosy employer might want to know whether their employees use their (private) phone when at work or a marketing company might be specifically interested in targeting users with excessive screen time.
Figure~\ref{fig:whatsapp-ios-screen-on-off} compares the observed RTT for WhatsApp with an active and inactive screen state on the iPhone.
An inactive screen leads to RTTs of about two seconds,
an active of about one second.
Extracting the screen time only worked with less frequent probing, e.g., one ping per 20 seconds as more frequent probes would have prevented the phone from pivoting into a deep sleep state.

\subsubsection*{Showcase II: Deriving IM Application Activity}

Further, we show that even the use of an application could be extracted from the measured RTTs.
In particular, we checked whether RTTs change if the target application is open, i.e., in foreground.
An envious ex might be curious about how much time their former partner spends on the messaging app potentially chatting with new acquaintances.
Figure~\ref{fig:whatsapp-ios-app-active-inactive} shows that the RTT drops to about 300\mseconds as soon as the application is opened on the phone.
If the application is closed, i.e., moved to the background, the timings dwell in an intermediary state (RTT: 500\mseconds) for about 30 seconds before normalizing at their initial level (RTT: 1\seconds) as the screen is still active.
To further support our findings, we looked for evidence of this behavior in the iPhone's system log (via \textit{idevicesyslog}). We are able to confirm that the application is first put on hold leading to the observed intermediary state for 30 seconds before being moved into standby.

\subsubsection{Behavior Fingerprinting on Android Devices}

For Android, we likewise found characteristic patterns allowing to differentiate between screen and application activity states on a variety of phones.
Due to the more diverse landscape of manufacturers, chipsets, OS flavors, and software versions, the patterns for certain activity states differ among models and need to be individually adjusted for each target device.
Yet, some general rules hold across all tested device models, e.g., deeper standby states causing increased jitter and higher RTTs.

\begin{figure}[t]
  \centering
  \includegraphics[width=.8\linewidth,trim={2.5mm 2.5mm 2.5mm 2.5mm},clip]{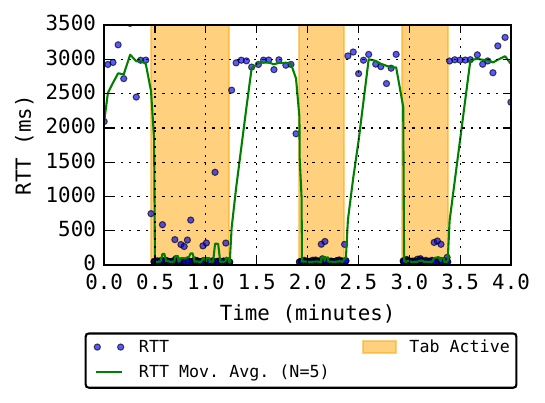}
  \caption{WhatsApp Web on Firefox (Windows): RTTs are 50\mseconds for an open browser tab and 3\seconds if the user switches to another tab. %
  }
  \label{fig:whatsapp-firefox}
\end{figure}

\begin{figure*}
    \centering
    \includegraphics[width=0.79\linewidth]{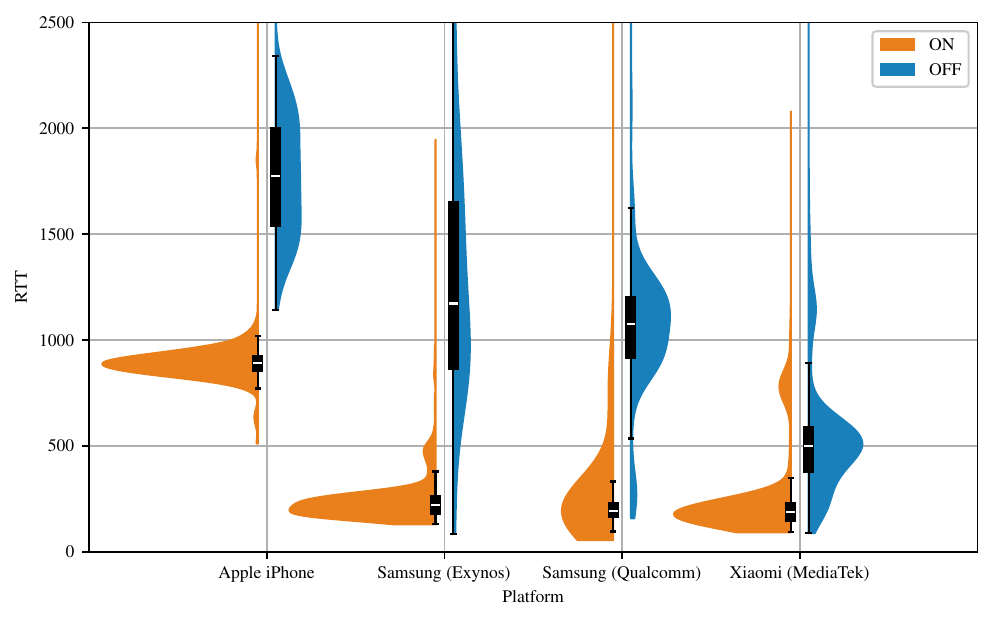}
    \\{\small Device models: Apple iPhone 11, Samsung Galaxy A54 5G, Samsung Galaxy S23, Xiaomi Poco M5s} \\
    \caption{Characteristic screen on/off timings for different manufacturers and chipsets (all measured as \emph{\creepy} over Wi-Fi with 1 ping per minute).
    The box and violin plots show, that we observed differences between timings for the screen on vs. screen off state across all tested devices. While this only shows the overall RTT distribution, local patterns (e.g., jitter) could be used to further refine the distinction between states.
    The leftmost graph (iPhone) corresponds to the data that is shown in Figure~\ref{fig:density-graph-iphone}.
    }
    \label{fig:android-screen-on-off-device-comparison}
\end{figure*}

Besides identifying characteristic patterns, 
selecting an appropriate ping frequency greatly influences a measurement's outcome and thus needs to be fine-tuned for specific phone models.
For example, on Samsung models, lower ping frequencies (e.g., 1 ping per minute) allow the phone to enter a deep sleep state, resulting in more distinct RTT differences between activity states.
Conversely, on Qualcomm- and MediaTek-based Xiaomi phones, higher probing frequencies (e.g., 1 ping per second) do not disrupt the phone’s standby behavior and still allow for a clear separation between active and inactive states.
To compare characteristic screen on/off timings for various manufacturers and models, we measured the RTTs for a range of smartphones.
Using a probing interval of one ping per second (to also cover standby states on Samsung models), we recorded delivery receipt RTTs for different screen states on each device.
For each individual phone and state, we made sure to gather at least 300 data points (i.e., 5 hours of capture time).
Figure~\ref{fig:android-screen-on-off-device-comparison} compares the RTT distribution for each screen state on our testing devices.
Despite distinct characteristics across models, differentiating between screen-on and screen-off timings appears feasible for all devices.
Two detailed plots that show characteristic patterns on Android phones can be found in Appendix~\hyperref[section:appendix-android-patterns]{A4}.

\subsubsection{Monitoring Behavior on Companion Devices}
\label{subsubsec:companiandevices}
Besides examining characteristic RTT patterns on main devices, we analyzed the RTTs on desktop- and web-companion devices.
We tried to differentiate between the active (i.e., application or corresponding browser tab in foreground) and inactive (i.e., application minimized or tab in background) state.

Figure~\ref{fig:whatsapp-firefox} shows that it is trivial to differentiate by RTTs between an active and inactive (i.e., browser tab46904690 in the background on Firefox) WhatsApp Web session.
While we got immediate responses (roughly within 50\mseconds) in the active state, responses took about 3s when another tab was focused or when the browser was minimized.
Moreover, the high response times occurred as soon as another window was fully covering the canvas of the Firefox window (i.e., the supposed standby mechanism kicks in as soon as the WhatsApp window is not actively painted on the screen).
Clearly, this behavior allows sophisticated tracking of the victim's Whatsapp usage within their browser.
For Firefox, all tested Operating Systems (Windows, Linux, Mac) showed the same behavior.
For the residual companion devices (other browsers and Desktop Apps), we did not see any obvious distinctions.

\begin{table}[t]
\centering
\begin{adjustbox}{max width=\columnwidth}
\begin{tabular}{ll l l}
\toprule
& \textbf{OS} & \textbf{Delivery Receipts} & \textbf{Read Receipts}  \\ \midrule
\parbox[t]{2mm}{\multirow{5}{*}{\rotatebox[origin=c]{90}{WhatsApp}}}
& Android & Separate           & Stacked            \\
& iOS     & Separate           & Stacked (Reversed) \\
& Web     & Stacked            & Stacked            \\
& Windows & Stacked            & Stacked            \\
& macOS   & Stacked (Reversed) & Stacked (Reversed) \\
\rule{0pt}{3ex}
\parbox[t]{2mm}{\multirow{3}{*}{\rotatebox[origin=c]{90}{Signal}}}
& Android & Separate           & Stacked            \\
& iOS     & Separate           & Stacked (Random) \\
& Desktop & Stacked            & Stacked (Reversed) \\ %
\bottomrule
\end{tabular}
\end{adjustbox}
\caption{Deivce/OS Fingerprinting: WhatsApp and Signal show different receipt handling for different platforms.}
\label{table:receipt-handling-whatsapp}
\end{table}
\begin{figure*}
    \centering
    \includegraphics[width=0.9\linewidth]{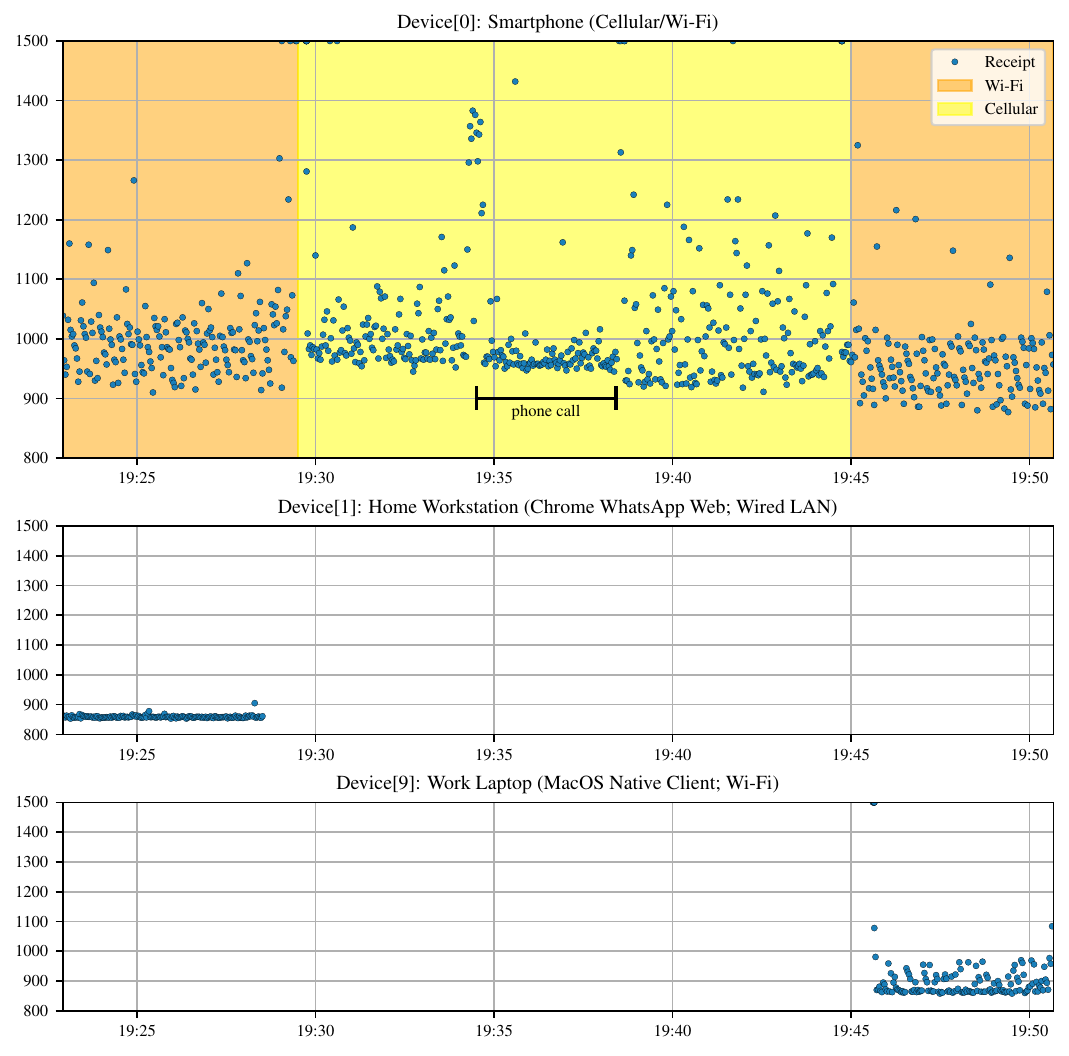}
    \caption{Real-World Tracking Scenario of two companion devices (web-client and native client) and the main device (smartphone) across different access technologies (LTE, Wi-Fi and Wired LAN) and usage scenarios (active usage, voice call, screen off phases), measured by a \emph{\spooky} with one ping every 2\seconds.}
    \label{fig:realworld}
\end{figure*}

\subsection{Device OS Fingerprinting}
\label{sec:device-os-fingerprinting}
Section~\ref{subsec:fingerprinting-user-behaviour} showed differences in RTTs that might be abused to observe a victim's behavior,
but we also noticed differences in the applications' implementations for different target architectures.
This includes handling delivery- and read receipts, e.g., receipt ordering and receipt stacking, for missed messages due to the device being offline.
Therefore, we sent multiple messages to an offline target.
As soon as the target device went online again, 
the pending messages were fetched from the server and acknowledged via delivery receipts.
As soon as the target opened the conversation, it sent out the corresponding read receipts.
Although the protocol supports stacking, i.e., combining the receipt of multiple messages into a list within one single receipt, 
some implementations always issue separate receipts.
Whenever summarized receipts were used, 
the messages' order was found to differ among different implementations (natural order vs. reversed vs. randomized order).
Table~\ref{table:receipt-handling-whatsapp} summarizes the found discrepancies.
While read receipts are only sent to \emph{\creepys} during an active conversation, the behavior regarding delivery receipts can also be measured by \emph{\spookys}.
These differences can be viewed as information disclosure that can be used to fingerprint a victim's system and to refine further exploits against the target.

\subsection{Real World Tracking Example}
\label{subsec:real-world-tracking}

While the previous examples clearly demonstrate the feasibility of extracting privacy-sensitive information, they were measured within a relatively stable and controlled environment (i.e., within testing devices in our lab).
We further show that attacking a user's privacy is also feasible in an open-world scenario even with limited prior information and thus created a measurement under real-world settings, monitoring the phone (a Xiaomi Poco X3 NFC device) of a volunteering colleague on their way to our lab.
The volunteer afterward provided us with information about their devices and their actions during the capture, which was used to annotate the corresponding Figure~\ref{fig:realworld}.
The sending device did not have any prior relation to the observed device, i.e., this measurement reflects the worst attack scenario of a \emph{\spooky}.

Before we started the measurement, we inspected our victim's device list (cf. Section~\ref{subsec:tracking-users-across-devices}): \texttt{[0, 1, 9]}. We can see that they currently have three existing sessions, one main device (index 0), and two companion devices (index 1, 9).
Due to the (auto-incrementing) index, we know in which order the sessions were created, i.e., the first companion device corresponds to a relatively old and stable device, while device 9 is a newer (or potentially just temporary) session.

After starting our probing, we receive receipts from two --currently online-- devices (cf. Section~\ref{subsec:monitoring-device-online}): the main device and companion device 1 (our volunteer's desktop computer, running WhatsApp Web in the browser).
According to our volunteer, device 1 is connected via LAN, which is reflected in the graph by very stable RTT timings with low jitter.
At 19:28, our volunteer turned off their desktop computer (thus, no more receipts are received from this device), shortly before leaving their flat and heading to the office.
The phone switched from Wi-Fi to LTE, which is reflected by a changed RTT pattern (slightly higher, but more dense RTT timings).
On their walk towards the office, the victim issues a phone call, again resulting in a more dense receipt distribution (since the phone is in a high activity state).
Finally, at 19:45, our victim arrives at the office, as their phone switches back from LTE to Wi-Fi.
Shortly after, our volunteer turned on their work laptop (device 9), which synchronizes all missed messages, causing it to send a lot of receipts for the previous probing requests.
Due to the reversed ordering of the stacked delivery receipts (cf. Section~\ref{sec:device-os-fingerprinting}), we know that this device is a macOS computer running the WhatsApp desktop client.
Comparing the jitter and density of the RTT patterns of the two companion devices (devices 1 and 9) we can easily spot the difference between LAN and Wi-Fi.

\subsection{Resource Exhaustion Attacks}
\label{subsec:resource-exhaustion}
Although covert messages are not displayed on the target's phone, they still use resources (e.g., traffic, battery, phone storage).
To amplify the resource exhaustion that can be achieved with a single message, we attempted sending different message actions with different payload sizes, aiming to detect server- and client-side limits.
Table~\ref{tab:max-message-sizes} shows the discovered size limits.
Interestingly, WhatsApp uses different boundaries, depending on the message size, and allows message reactions to carry up to 1\,MB of payload data.
While the client-side limit of actually handling the message seems to be much lower (i.e., no delivery receipts are issued for reactions containing more than 30 bytes of data), the message is still received and processed before it is discarded.

\begin{table}[]
\centering
\begin{tabular}{@{}lrrrrr@{}}
\toprule
           &  \textbf{Send}     & \textbf{Edit}     & \textbf{React}     & \textbf{Delete} & \textbf{Consumable Data} \\ \midrule
\iWhatsapp & 65       & 65       & 1,000     & -      & 13,320\,MB/h    \\
\iSignal   & 194      & 194      & 194       & -      &    360\,MB/h    \\ \bottomrule
\end{tabular}
\caption{Server-side payload limits (in KB) for different message types.
(Invalid) reactions can contain arbitrary data and are not displayed at the target.
Thus, besides abusing them for resource exhaustion, they could also be utilized as a covert channel or for data exfiltration.}
\label{tab:max-message-sizes}
\end{table}

\parvspace
\noindent\textbf{Traffic Inflation.}
Our traffic inflation measurements for \mbox{WhatsApp} showed that an attacker can cause 3.7\,MB per second (i.e., 13.3\,GB per hour)\footnote{Capture period: 2 hours. In addition to our captured traffic dump, both the phone's system-level data usage statistics and WhatsApp's internal data consumption view confirmed the volume of traffic generated.} of data traffic for the victim without the latter receiving any notification in the application.
This value was reached by a single client session continuously sending message reactions with 1\,MB of payload and might be further amplified using multiple clients or sessions.
The attack covertly inflates a victim's data bill and might use the bandwidth planned for other applications, potentially leading to their denial of service.

\parvspace
\noindent\textbf{Battery Drainage.}
Besides using up a user's data allowance, receiving many and large messages additionally drains the smartphone's battery.
We measured the battery exhaustion on three phones by blasting large reaction messages via WhatsApp for a period of one hour.
While regular (idle) battery drainage for all phones was less than 1\,\% per hour, we were able to drain a considerable share of the battery (iPhone 13 Pro: 14\,\% per hour, iPhone 11: 18\,\% per hour, Samsung Galaxy S23: 15\,\% per hour)\footnote{Again, the system's battery usage overview confirmed that WhatsApp was responsible for the observed battery drain.}.
During our tests the phones were on normal standby (screen off), connected to Wi-Fi and all attacks were executed by \emph{spooky strangers}.
For Signal, we were not able to considerably drain the battery of our testing phone (iPhone 13 Pro).
Due to considerably stricter rate limits, it only decreased by 1\,\% after an hour of attack.

\section{Related Work}

\textbf{Mobile Instant Messaging Security and Privacy.}
Instant messenger security has been investigated since their early days.
Schrittwieser et al.~\cite{schrittwieser_guess_2012} analyzed attack vectors exploiting insufficient authentication in nine messenger applications.
Back then, WhatsApp was found to be vulnerable to account hijacking, the unauthorized modification of the users' status pages,
the unlimited delivery of unrequested SMS, and user account enumeration.
A subset of attacks continued to exist over multiple years~\cite{mueller_what_2014,hagen_all_2021, hagen_contact_2022, gegenhuber_2025_heythere}
and user account enumeration even become feasible for the more security-oriented messenger Signal~\cite{hagen_all_2021, hagen_contact_2022}.
With the advent of E2EE, Be'ery~\cite{beery_whatsapp_2024} and Gegenhuber et al.~\cite{gegenhuber_2025_prekeypogo} discussed undesirable leaks of multi-device architectures,
relevant for WhatsApp.
For E2EE, one public key has to be maintained per device in the application's inventory and user behavior (i.e., addition or change of device) might be observed by changes in an account's public keys.
Based on the delivering session, a receiver is also able to infer the sender's device issuing a message.
Our research emphasizes that this also holds for Signal.
Beyond, we discovered further privacy implications by exploiting delivery receipts as a side channel for WhatsApp and Signal.

\parvspace
\noindent\textbf{Delivery and Read Receipts.}
Instant messengers typically acknowledge receipt and reading of a message.
Reading receipts facilitate stalking,
e.g., in the context of intimate partner abuse~\cite{freed_stalker_2018},
even by weak adversaries that are bound to the messenger's regular user interface.
It is now possible to disable these reading receipts,
but delivery receipts are continuously returned by WhatsApp, Signal and Threema.
Simulating a regular WhatsApp conversation with the victim,
delivery receipts were used to narrow down user location (e.g., UAE vs Germany) and 
to distinguish between cellular- and Wi-Fi-based connections~\cite{schnitzler_hope_2023}.
Thereby, the adversary requires an on-going conversation with the victim and each probe triggers another message,
i.e., the attack remains overt to the victim.
The principal idea has been transferred to cellular networks~\cite{bitsikas_freaky_2023},
and later extended by multi-location measurements to improve accuracy~\cite{bitsikas_amplifying_2024}.
The latter approaches rely on delivery reports of silent SMSes. 
Thereby, the victim remains unaware of the attack,
enabling geolocation also at unusual time of the day or at regular intervals.
In our work,
we show how to trigger delivery receipts of instant messengers without any notification of the victim and in the absence of an on-going conversation,
rendering WhatsApp- and Signal-based attacks as stealthy as silent SMS-based ones.
Overall, the impact on privacy is more substantial
as we are not only able to infer a user's geolocation
but also more detailed information on device activity and user behavior (e.g., screen on/off, browser active/inactive).

\parvspace
\noindent\textbf{Battery Drain of Instant Messengers.}
Battery draining attacks had already been known in the era of feature phones~\cite{racic_exploiting_2006}
exploiting MMS services to consume battery up to 22 times faster,
and only later transferred to smart phones~\cite{moyers_effects_2010}.
Regular battery drain of instant messengers like WhatsApp has been investigated as early as 2014,
proposing message bundling to save energy~\cite{vergara_mice_2014}.
More recent work on mobile Tor use, points into a similar direction~\cite{hugenroth_powering_2023}. 
Consumption is primarily caused by radio transmission and might be reduced by adequate message scheduling,
whereas consumption due to cryptography is negligible.
The significance of our battery drain attacks lies in its versatility and stealth.
Two billion WhatsApp users, i.e., a fourth of the world population, might become a victim to our attack,
again an on-going conversation with the adversary is not a prerequisite.

\parvspace
\noindent\textbf{Covert Channels.}
Camoufler uses the Signal infrastructure as a tunnel to evade Internet censorship~\cite{sharma_camoufler_2021} 
as censors fear collateral damage caused by the prohibition of popular messaging applications.
Our covert channels follow a different rationale.
Instead of disguising traffic from a censor,
they evade visible representation in the messenger's user interface.

\parvspace
\noindent\textbf{Security and Privacy Issues in Cellular Services.} %
Beyond Over-the-Top (OTT) applications like WhatsApp and Signal, prior work has demonstrated that tracking a mobile user's geolocation (e.g., under roaming conditions) is possible via the traditional cellular network~\cite{gegenhuber_2023_mobileatlas, bitsikas_freaky_2023, bitsikas_amplifying_2024}.
3GPP-standardized messaging services that are terminated over third-party Internet connections, such as VoWiFi and RCS, have also been shown to be vulnerable to various security and privacy issues~\cite{tu_2016_new, gegenhuber_2024_diffie, gegenhuber_2024_geoblocking, yang_2024_uncovering}.
In contrast, the present work does not rely on native 3GPP services but instead requires an additional OTT application to be installed on the phone.

\section{Mitigations}
\label{sec:mitigation}
    \parvspace
    \noindent\textbf{Restricting Delivery Receipts.}
    Our measurements show that all three analyzed messengers also send delivery receipts for unknown users that are not in the victim's contact list.
    Restricting this feature to real conversations and automatically dropping messages or preventing receipts for unknown numbers would hinder \emph{\spookys} from tracking arbitrary victims.
    Additionally, privacy-conscious users should be able to disable the instant transmission of delivery receipts.

    \parvspace
    \noindent\textbf{Coarser Receipt Timings.}
    Letting the sender know that a message was successfully received can be a convenient feature in an asynchronous conversation.
    However, there are no strict real-time requirements, i.e. the perceived experience does not change when this information is only updated after a few seconds.
    Adding noise to these acknowledgment timings would easily prevent tracking (i.e., geolocation- and activity monitoring) based on the receipt's RTTs.

    \parvspace
    \noindent\textbf{Improve Client-side Validation.}
    When messages are not E2EE, they can be validated by the server and only forwarded to the receiver when passing the validation.
    However, this server-side validation is not possible with E2EE, requiring more rigorous validation by the receiving client.
    For example, many of the presented attacks are not possible when clients properly validate the referenced message IDs and thus discard invalid messages (instead of acknowledging them via a delivery receipt).
    While our primary focus is on privacy-related issues, the shift from server-validated input to E2EE content is particularly important from a security standpoint.
    Parsing unvalidated data can quickly introduce severe security vulnerabilities.

    \parvspace
    \noindent\textbf{Rate Limiting.}
    In our measurements, we were able to drain a user's data quota and battery by sending large messages over a prolonged time.
    In contrast, regular (text) messaging only needs very limited bandwidth (note: media messages are usually transmitted over separate media servers).
    Thus, employing restrictive messaging rate limits on the server side could mitigate these attacks.
    Moreover, receiving an excessive amount of messages could also be automatically detected by the receiver and then trigger a UI notification and (temporarily) block the corresponding phone number.

    \parvspace
    \noindent\textbf{Syncronized Multi-Clients.}
    To cope with multi-device leakage, devices could prioritize synchronizing their state before issuing receipts for recent messages. While only introducing a minor timing overhead, this would ensure that a delivery receipt is just sent once. Alternatively, other proposals for multi-device protocols consider hiding the amount of companion devices~\cite{campion_multi-device_2019}.
    
    \parvspace
    \noindent\textbf{Harmonizing Client Behavior.}
    Supporting different operating systems often requires having multiple codebases that are written in different programming languages.
    In many cases, these different implementations behaved inconsistently in how they responded to specific messages, introducing fingerprinting possibilities for the attacker.
    Harmonizing client behavior or moving towards a single code base that can be used across different platforms could solve this problem.

\section{Discussion}

Delivery receipts have already been known as veritable timing side channels compromising user privacy.
In this work,
we discovered the existence of stealth delivery receipts in two major messaging services -- market leader WhatsApp and its security-oriented alternative Signal.
This way, an adversary is able to trigger delivery receipts at another user's client without leaving a trace for the latter --
indeed, stealth probing makes long-term and high-frequency probing only possible in the first place and surpasses previous possibilities of remote observation by far.
State-of-the-art E2EEE encryption, requiring an individual encryption key per registered device, only exacerbates the situation as it enables the attribution of messages to a user's different devices (mobile phone, desktop, web).
This way,
we are able to remotely create comprehensive observation profiles of victims solely by observing delivery receipts.
Beyond that, we are able to stealthily launch resource exhaustion attacks (data quota, battery) against mobile phones.

Our attacks' impact is significant:
First, the requirements for the attacker are low.
They only need a phone number for registration  at the messaging service and a mobile phone --
a prepaid card and an older phone model are perfectly adequate.
Second, with more than two billion users on WhatsApp, the number of potential victims is vast.
According to our results,
Signal -- dedicatedly developed with security and privacy in mind -- does not appear to protect its users better and might even put high-profile users like US Senate~\cite{whittaker_encryption_2024} and European Commission~\cite{cerulus_eu_2020} staff at risk.
Moreover, recent media revelations have shown that high-ranking US officials, including the Secretary of Defense, use both Signal and WhatsApp for personal and professional communication~\cite{atlantic_2025_houthi}, and in some cases even have their phone numbers publicly accessible online~\cite{spiegel_private_2025}, making them easy targets for such attacks.
The quality of the results varies somewhat between different phones,
but our results pinpoint that practically all of them are affected.
But even when only looking at the iPhone with its particular clear measurement results,
its worldwide market share of 20 to 30\,\% renders several hundred million people vulnerable.

From a user perspective, the situation is particularly dire as an individual cannot take any protective measures except the complete deinstallation of the service.
Due to the attacks' stealth, they do not realize an ongoing attack either (with the obvious exceptions of drained battery and overdrawn quota as a consequence of resource exhaustion attacks).
Moreover, users are unable to locate the source of the attack -- it is important to note that the adversary does not need to be in the victim's contact list.
It is therefore essential that the operators of the messaging server take action and implement security measures like those suggested in Section~\ref{sec:mitigation}.

\subsection{Ethical Considerations}
In the course of our research, we only probed WhatsApp, Signal, and Threema accounts belonging to and used by the authors of this paper, all of whom provided their explicit consent.
Most of the accounts used in our experiments were test accounts created specifically for the purpose of these measurements.
The devices used to explore and demonstrate the impact of various usage and environmental scenarios were either i)~dedicated lab devices intended solely for research purposes, or ii)~personal devices owned by the test operator (i.e., one of the authors).
The volunteer involved in the real-world experiment described in Section~\ref{subsec:real-world-tracking} was also one of the authors, and was therefore fully informed about the associated security and privacy implications.
Tracking was conducted strictly within the agreed-upon duration of the experiment.

\parvspace
\noindent\textbf{Server Infrastructure.}
As the traffic is E2EE, the messenger infrastructure remains unaware of and thus unaffected by non-compliant messages, e.g., the deletion/modification or reaction to nonexistent messages.
The application servers only see the overall traffic volume and pattern that is forwarded from our sending to the receiving accounts.
We assume that the messengers' infrastructure is laid out for massive data forwarding,
and consider our maximum rate of 3.7\,MB/s, as caused by the resource exhaustion attacks, a moderate load for an infrastructure serving more than two billion users.

\parvspace
\noindent\textbf{Responsible Disclosure.}
Our attacks heavily inflict user privacy and additionally enable resource exhaustion (data quota, battery).
Beyond, they affect all ($>$ 3 billion) users of the messenger platforms WhatsApp and Signal.
For both messenger operators, we consequently submitted our findings to their security contacts on September 5th, 2024. 
On September 24th, 2024, we received a confirmation receipt from Meta, responsible for WhatsApp, indicating that our results had been passed on to the relevant development team but have not received a substantive response ever since.
On August 8th, 2025, more than 11 months after the initial report, we once again received the information that the report was reviewed by the security team and has been forwarded to the relevant engineering team.

As of November 14th, 2024, it appears that the Firefox activity leakage as described in Section~\ref{subsubsec:companiandevices} has been fixed but we did not receive any more detailed information.

From the Signal Technology Foundation, we did not get any answer at all.

\parvspace
\noindent\textbf{Open Science.}
We believe in open science and therefore plan to release our modified clients in the future.
However, we will not release them at the current publication date (2025-08-14), as the issues we identified have not yet been addressed by the platform operators.
Through the responsible disclosure process, we expect these vulnerabilities to be resolved eventually, ensuring that our modified clients will no longer pose any risk of misuse.

\section{Conclusion}
In this work, we demonstrated that modern E2EE messaging architectures like WhatsApp and Signal unintentionally expose privacy-sensitive information about their users. Specifically, an adversary armed with only a target's phone number can determine the exact amount, type, and online status of the target's devices. Furthermore, detailed behavioral patterns, such as screen time or messaging app usage duration, can be inferred with a resolution down to the second.

This vulnerability is exploited through covert probing messages that trigger delivery receipts without generating any notification within the targeted application, akin to a stealth SMS. Additionally, the structure of E2EE messaging, combined with the absence of server-side message quotas, enables attackers to misuse these capabilities for resource exhaustion attacks draining a target's battery or data allowance. Notably, there is currently hardly anything a targeted user can do about this for multiple reasons. These attacks neither cause any notification on the targeted device, nor require an active conversation between the attacker and the target, nor can the attacking account be blocked or reported, nor is the deactivation of delivery receipts entirely possible at the moment.

Our findings reveal that %
mechanisms embedded in modern E2EE messaging architectures -- such as delivery receipts and multi-device support -- can have significant implications on user privacy. Consequently, it is essential to balance functional requirements, usability and convenience with privacy and security, particularly in E2EE applications that are inherently privacy-sensitive per design.

\section*{Acknowledgment}
This material is based upon work partially supported by
(1) the University of Vienna, Faculty of Computer Science, Security \& Privacy Group,
(2) the University of Vienna, Faculty of Computer Science, Communication Technologies Group,
(3) the FFG Bridge project 46322124 SecKey, 
(4) the FFG KIRAS/K-PASS project 59103683 TelCrit,
(5) the Austrian Science Fund (FWF) (SFB SPyCoDe F85),
(6) SBA Research (SBA-K1 NGC) is a COMET Center within the COMET – Competence Centers for Excellent Technologies Programme and funded by BMIMI, BMWET, and the federal state of Vienna. The COMET Programme is managed by FFG.

\bibliographystyle{plain}
\bibliography{citations,background}

\appendix

\section{Appendix}

\gge{IEEE SP Submission Guidlines: Submitted papers may include up to 13 pages of text and up to 5 pages for references and appendices, totaling no more than 18 pages. All text and figures past page 13 must be clearly marked as part of the appendix. The final camera-ready paper must be no more than 18 pages, although, at the PC chairs’ discretion, additional pages may be allowed. Reviewers are not required to read appendices.}

\subsection{Messenger Infrastructure Analysis}
\label{sec:infra}
As a basis for our research,
we investigated the messaging services' infrastructure for submitting and receiving messages and provide novel insights.
Applying source code analysis and inspecting real-world network traffic of our mobile devices, 
we discovered the relevant web endpoints and domains, see Table~\ref{tab:overview-api-domains}, in a first step.

\begin{table}[h]
    \centering
    \begin{adjustbox}{max width=\columnwidth}
    \begin{tabular}{@{}ll@{}}
    \toprule
    \textbf{App} & \textbf{Domain}                                 \\ \midrule
    \iWhatsapp     & web.whatsapp.com (web), g.whatsapp.net (mobile) \\
    \iSignal       & chat.signal.org                                 \\
    \iThreema      & ds.g-xx.0.threema.ch                            \\ \bottomrule
    \end{tabular}
    \end{adjustbox}
    \caption{Domain names that are used to connect to the messaging services (all available via dual-stack). Usually a websocket connection (port 443) is used. The only exception are mobile WhatsApp clients connecting directly via port 5222 (xmpp).}
    \label{tab:overview-api-domains}
\end{table}

In a second step, we investigated these endpoints from different vantage points in the AWS cloud in a threefold manner:
\begin{itemize}
    \item \roundframe{L1} DNS resolution of the endpoint domains, 
    \item \roundframe{L2} measurement of the application-agnostic latency by probing with ICMP and TCP, 
    \item \roundframe{L3} measurement of application-level latency between a messaging client and the server with the applications' \\keepalive/heartbeat functionality.
\end{itemize}

While \roundframe{L1} and \roundframe{L2} is feasible for any domain name, 
\roundframe{L3} requires client emulation and thus use and adaptation of the software projects presented in Table~\ref{tab:overview-api-access}.

\parvspace
\noindent
\subsubsection*{\iWhatsapp~WhatsApp}
WhatsApp uses GeoDNS to route traffic from clients, based on their source IP address, to different target addresses,
see also previous work on these aspects~\cite{schnitzler_hope_2023}.
A reverse DNS lookup of the latter addresses reveal their domain names, containing the three letter airport code of the edge locations (e.g., 
\texttt{whatsapp-cdn-shv-01-vie1.fbcdn.net}\sloppy for Vienna, Austria). 
Ensured by GeoDNS, client and edge locations are close, leading to \roundframe{L2} latencies of 1 - 10\mseconds when measured from our AWS instances.

Our measurements for \roundframe{L3} latencies measured with \mbox{WhatsApp} keepalives are however significantly higher.
Further analysis showed, that the edge location revealed by DNS only plays a minor role for the overall messaging RTTs, since they only serve as an entry node to Meta's internal network.

In fact, message delivery is handled via a lower number of centralized messaging servers.
We discovered eight such servers (represented by a three letter location attribute), three within Europe and six in the US:
\begin{itemize}
\item \textbf{\texttt{odn}}: Odense, Denmark
\item \textbf{\texttt{cln}}: Clonee, Ireland
\item \textbf{\texttt{lla}}: Luleå, Sweden
\item \textbf{\texttt{frc}}: Forest City, North Carolina, US
\item \textbf{\texttt{atn}}: Altoona, Iowa, US
\item \textbf{\texttt{nao}}: New Albany, Ohio, US
\item \textbf{\texttt{rva}}: Sandston, Virginia, US
\item \textbf{\texttt{vll}}: Huntsville, Alabama, US
\item \textbf{\texttt{cco}}: Prineville, Oregon, US (same as \texttt{prn}) %
\end{itemize}

The selection of the central messaging server is influenced by the client's \verb|routing_info| cookie.
If it is empty, as for example in the very first connection attempt, a random messaging server is assigned.
Upon re-connection, the client communicates its cookie in the connection handshake, indicating the previously used (and usually closest) location, effectively pinning the proposed server.

In conclusion, a message between two messaging clients that are connected to the same edge location cannot be directly forwarded, but rather takes a detour via the clients' messaging servers.

\parvspace
\noindent
\subsubsection*{\iSignal~Signal}
Signal uses a set of static IP addresses for all messenger clients (\roundframe{L1}).
Pinging these IP addresses (\roundframe{L2}) from our AWS instances results in RTT of less than 1\mseconds from all locations suggesting the use of Anycast routing.
A reverse DNS lookup reveals that these IP addresses belong to AWS Global Accelerator\footnote{\href{https://aws.amazon.com/global-accelerator/features}{\texttt{aws.amazon.com/global-accelerator/features}}}, 
a service providing a static entry to applications hosted within the Amazon cloud.
Measuring the \roundframe{L3} RTT from all 29 AWS EC2 regions suggests that the service is hosted within \mbox{\textit{us-east-1}}.
Measuring from there, we see an application-layer keepalive RTT of only 3\mseconds.
In comparison, the corresponding RTT from \textit{us-west-1} is 62\mseconds and goes up to 252\mseconds for \textit{ap-southeast-3}.

\parvspace
\noindent
\subsubsection*{\iThreema~Threema}
Threema uses static unicast IP addresses for all messenger clients (\roundframe{L1}).
According to their website\footnote{\href{https://threema.ch/en/faq/server_location}{\texttt{threema.ch/en/faq/server\_location}}}, 
they host their infrastructure in Zurich, Switzerland.
Our plausibility checks confirm this as we see low RTTs for measurements from vantage points in Central Europe (\roundframe{L2}:~5\mseconds, \roundframe{L3}:~50\mseconds) and increased RTT from the US (e.g., \roundframe{L2}:~90\mseconds, \roundframe{L3}:~140\mseconds).

\subsection{Testing Devices}
\label{sec:testingdevices}
We conducted our tests on seven devices from three different vendors, of which two run iOS and five Android.
To maximize the validity of our results, we included devices from the top three global manufacturers for both smartphone brands (Apple, Samsung, Xiaomi) and modem chipsets (MediaTek, Qualcomm, Exynos).
All devices operated on up-to-date software, including both the underlying OS and target applications.

\begin{table*}[t]
    \centering
    \begin{tabular}{@{}lllll@{}}
    \toprule
    Device                & Modem Chipset & OS                       & WhatsApp   & Signal \\ \midrule
    iPhone 13 Pro         & Qualcomm      & iOS 17.6.1               & 2.24.17.78 & 7.26   \\
    iPhone 11             & Intel         & iOS 17.6.1               & 2.24.17.78 & 7.26   \\
    Samsung Galaxy S23    & Qualcomm      & Android 14               & 2.24.17.79 & 7.15.4 \\
    Samsung Galaxy A54 5G & Exynos        & Android 14               & 2.24.17.79 & 7.15.4 \\
    Xiaomi Poco M5s       & MediaTek      & Android 13 (MIUI 14.0.4) & 2.24.16.10 & 7.15.4 \\
    Xiaomi Poco M3 Pro 5G & MediaTek      & Android 13 (MIUI 14.0.4) & 2.24.16.10 & 7.15.4 \\
    Xiaomi POCO X3 NFC    & Qualcomm      & Android 12 (MIUI 14.0.5) & 2.24.16.10 & 7.15.4 \\
    \bottomrule
    \end{tabular}
    \caption{Overview of the devices including software versions that were used throughout our tests.}
    \label{tab:testing-devices}
\end{table*}

\subsection{Geolocation with Delivery Receipts}
\definecolor{redline}{RGB}{184,84,80}
\definecolor{redfill}{RGB}{248,206,204}

\pgfdeclarelayer{background}
\pgfsetlayers{background,main}
\begin{figure}[tb]
\centering
\begin{adjustbox}{width=0.85\columnwidth}
\begin{tikzpicture}[
    sectionnode/.style={text width=3cm, text centered,font=\sffamily\small},
	squarednode/.style={rectangle, draw=gray!60, fill=gray!5, very thick, minimum size=8mm, text width=8.1em, text centered, font=\sffamily\small},
	roundednode/.style={rectangle, draw=gray!60, fill=gray!5, very thick, rounded corners, minimum size=5mm, font=\sffamily, text width=8em, text centered},
	attacknodered/.style={rectangle, draw=redline, fill=redfill, very thick,  minimum size=8mm, font=\sffamily\footnotesize, text width=8em, text centered},
	textnode/.style={minimum size=8mm, text width=6em, text centered},
	font=\sffamily
    ]
    \node[roundednode] (sender)  at (0,0) {(\textbf{S})ender};
    \node[roundednode] (server)   at (4,0) {(\textbf{M})essenger Server};
    \node[roundednode] (receiver)   at (8,0) {(\textbf{R})eceiver};
    
    \node[draw=none,fill=none] at (1,-1.1){\includegraphics{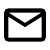}};
    \node[draw=none,fill=none] at (5,-1.2){\includegraphics{tikz/msg.pdf}};
    \node[draw=none,fill=none] at (3,-2.1){\includegraphics{tikz/check1.pdf}};
    \node[draw=none,fill=none] at (7,-3.4){\includegraphics{tikz/check1.pdf}};
    \node[draw=none,fill=none] at (3,-3.5){\includegraphics{tikz/check2.pdf}};
	\node[draw=none,fill=none] at (-0.6,-2.4){\includegraphics{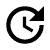}};
    
	\draw[-, very thick] (sender.south) -- (0,-4);
	\draw[-, very thick] (server.south) -- (4,-4);
	\draw[-, very thick] (receiver.south) -- (8,-4);
	
	\draw[->, very thick, dashed] (0,-1.5) -- (4,-1.5);
	\draw[->, very thick, dashed] (4,-1.6) -- (8,-1.6);
	\draw[->, very thick, dashed] (4,-1.7) -- (0,-1.7);
	
	\draw[->, very thick, dashed] (8,-3) -- (4,-3);
	\draw[->, very thick, dashed] (4,-3.1) -- (0,-3.1);

\end{tikzpicture}
\end{adjustbox}
\caption{Whats App Message Flow as discovered by Schnitzler et al.~\cite{schnitzler_hope_2023}. Our measurements show that these servers only serve as entry points to the Meta network.}
\label{fig:original-message-flow}
\end{figure}

\definecolor{redline}{RGB}{184,84,80}
\definecolor{redfill}{RGB}{248,206,204}

\pgfdeclarelayer{background}
\pgfsetlayers{background,main}
\begin{figure}[tb]
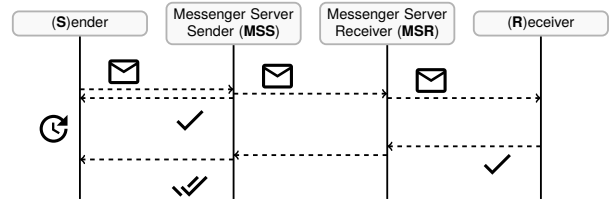

\centering
\begin{adjustbox}{width=\columnwidth}
\begin{tikzpicture}[
    sectionnode/.style={text width=3cm, text centered,font=\sffamily\small},
	squarednode/.style={rectangle, draw=gray!60, fill=gray!5, very thick, minimum size=8mm, text width=2em, text centered, font=\sffamily\small},
	roundednode/.style={rectangle, draw=gray!60, fill=gray!5, very thick, rounded corners, minimum size=2mm, font=\sffamily, text width=8em, text centered},
	attacknodered/.style={rectangle, draw=redline, fill=redfill, very thick,  minimum size=8mm, font=\sffamily\footnotesize, text width=8em, text centered},
	textnode/.style={minimum size=2mm, text width=2em, text centered},
	font=\sffamily
    ]
    \node[roundednode] (sender)  at (0,0) {(\textbf{S})ender};
    \node[roundednode] (server_sender)   at (3.5,0) { Messenger Server Sender (\textbf{MSS})};
    \node[roundednode] (server_receiver)   at (7,0) { Messenger Server Receiver (\textbf{MSR})};
    \node[roundednode] (receiver)   at (10.5,0) {(\textbf{R})eceiver};
    
    \node[draw=none,fill=none] at (-0.6,-2.4){\includegraphics{tikz/time.pdf}};
    \node[draw=none,fill=none] at (1,-1.1){\includegraphics{tikz/msg.pdf}};  
    \node[draw=none,fill=none] at (4.5,-1.2){\includegraphics{tikz/msg.pdf}};  
    \node[draw=none,fill=none] at (8,-1.3){\includegraphics{tikz/msg.pdf}}; 

    \node[draw=none,fill=none] at (2.5,-2.2){\includegraphics{tikz/check1.pdf}};
    \node[draw=none,fill=none] at (9.5,-3.2){\includegraphics{tikz/check1.pdf}}; 

    \node[draw=none,fill=none] at (2.5,-3.7){\includegraphics{tikz/check2.pdf}};

	\draw[-, very thick] (sender.south) -- (0,-4);
	\draw[-, very thick] (server_sender.south) -- (3.5,-4);
	\draw[-, very thick] (server_receiver.south) -- (7,-4);
    \draw[-, very thick] (receiver.south) -- (10.5,-4);
	
	\draw[->, very thick, dashed] (0,-1.5) -- (3.5,-1.5);   
	\draw[->, very thick, dashed] (3.5,-1.6) -- (7,-1.6);   
	\draw[->, very thick, dashed] (7,-1.7) -- (10.5,-1.7);  

	\draw[->, very thick, dashed] (7,-3) -- (3.5,-3); 
 	\draw[->, very thick, dashed] (10.5,-2.8) -- (7,-2.8);       

	\draw[->, very thick, dashed] (3.5,-3.1) -- (0,-3.1);
 	\draw[->, very thick, dashed] (3.5,-1.7) -- (0,-1.7);

\end{tikzpicture}
\end{adjustbox}
\caption{Updated WhatsApp Message Flow: Sender and receiver each connect to one of eight messenger servers. Messages are forwarded by both servers to reach the intended destination. 
}
\label{fig:extended-message-flow}
\end{figure}

Sending a message to a receiver triggers two acknowledgments, see Figure~\ref{fig:original-message-flow}.
First, the messenger server acknowledges the receipt of the message and forwards it to the receiver.
The receiver returns a delivery receipt to the server that is eventually forwarded to the sender.
Previous work~\cite{schnitzler_hope_2023} subtracted the RTT between message sending and receipt of the first acknowledgment 
from the total RTT to estimate the RTT between the server and the receiver. 
The latter is then used for coarse geolocation of smartphones (e.g., UAE vs. Germany).
Based on our insights on messaging infrastructure, all three messengers facilitate such coarse geolocation.
For Signal and Threema, an adversary is able to measure the RTT between the victim and the central server in Amazon's \mbox{\textit{us-east-1}} region and Zurich, respectively. 

Our infrastructure analysis however refines previous results~\cite{schnitzler_hope_2023} for WhatsApp,
see our updated message and delivery receipt flow in Figure~\ref{fig:extended-message-flow}.
First, the message is forwarded to the sender's messenger server triggering an acknowledgment;
then, forwarded to the receiver's messenger server before eventually reaching the receiver.
The latter then issues a delivery receipt that is forwarded by both servers before reaching the sender.
At first sight, WhatsApp, providing a total of eight such messaging servers appears to allow multilateration
i.e., the measurement from multiple vantage points for more precise geolocation.
This is however not true as the victim individually chooses its server via the \verb|routing_info| cookie.
The adversary is only able to discover this server by choosing the one with the lowest latency after iterating through all of them.
Once connected to the same server, the adversary is again able to conduct the same coarse-grained geolocation as with Signal or Threema.

\subsection{Characteristic Patterns for Android Phones}
\label{section:appendix-android-patterns}
\begin{figure}[h]
  \centering
  \includegraphics[width=.85\linewidth]{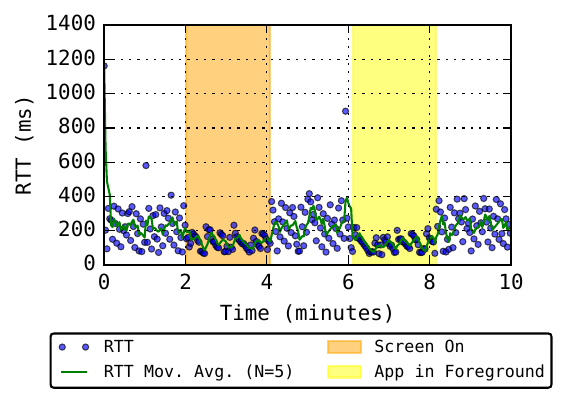}
  \caption{For the MediaTek-based Xiaomi Poco M3 Pro 5G, we used higher probing frequencies (one ping every 2\seconds) to differentiate between an active and inactive screen with second-level granularity (WhatsApp, \emph{\creepy}, Wi-Fi).}
  \label{fig:android-mediatek-xiaomi-screen}
\end{figure}

\begin{figure}[h]
  \centering
  \includegraphics[width=.85\linewidth]{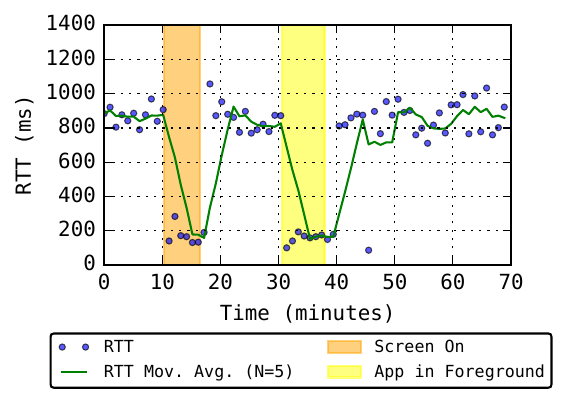}
  \caption{For the Samsung Galaxy S23, we needed to lower the probing frequency to one ping per minute, to be able to differentiate between an active and inactive screen (\mbox{WhatsApp}, \emph{\creepy}, Wi-Fi).}
  \label{fig:android-samsung-screen}
\end{figure}

On Android, different probing frequencies worked differently well, depending on the target phone model.
Two characteristic examples are shown in Figure~\ref{fig:android-mediatek-xiaomi-screen} and Figure~\ref{fig:android-samsung-screen}.

\subsection{iPhone}
\label{sec:appendix-iphone}
We systematically measured the RTTs within different environments (e.g., on an iPhone~13~Pro and iPhone~11, and via a cellular vs. LTE data connection) and applications (i.e., WhatsApp and Signal).
The methodology was switching between active and inactive phases. The first active phase corresponds to a normal phone unlock (e.g., with an active homescreen), while the second active phase corresponds to the IM application being in the foreground.
Each phase corresponds to 2 minutes for the measurements with one ping every 2 seconds and 10 minutes for one ping every 20 seconds respectively.

Figure~\ref{fig:iPhone-whatsapp-measurements} shows that for WhatsApp, the observed timings for \emph{\creepys} and \emph{\spookys} differ.
For Signal (cf. Figure~\ref{fig:iPhone-signal-measurements}) there is no such differentiation (i.e., probing as a \emph{\spooky} leads to the same RTTs as probing as a \emph{\creepy}).
Comparing the plots for WhatsApp and Signal, we see that some OS-specific patterns (e.g., application switch from foreground to background, as presented in Figure~\ref{fig:whatsapp-ios-app-active-inactive}) occur across both applications.
Note that, WhatsApp shows a counter-intuitive pattern for all four \emph{\spookys} cases (Figure~\ref{fig:artefact1} to Figure~\ref{fig:artefact2}), since the RTTs within the \textit{WhatsApp in Foreground} phase are consistently higher than in the \textit{Screen On} phase.
We verified that this is NOT a measurement error, by repeating the corresponding measurements multiple times.
Further analysis on systems level as well as access to WhatApp source code, could improve the accuracy of predicated usage patterns.

\newcommand\figwidth{0.39}
\begin{figure*}[h]
    \centering    
    \hfil
    \begin{subfigure}[b]{\figwidth\textwidth}
        \centering
        \includegraphics[width=\textwidth,trim={2.5mm 2.5mm 2.5mm 2.5mm},clip]{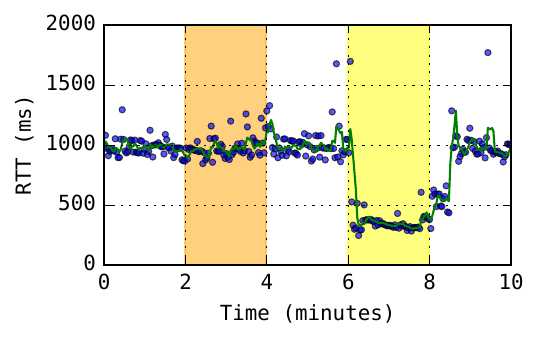}
        \caption{\emph{\Creepy}, 2\seconds interval, Wi-Fi}
    \end{subfigure}
    \hfil
    \begin{subfigure}[b]{\figwidth\textwidth}
        \centering
        \includegraphics[width=\textwidth,trim={2.5mm 2.5mm 2.5mm 2.5mm},clip]{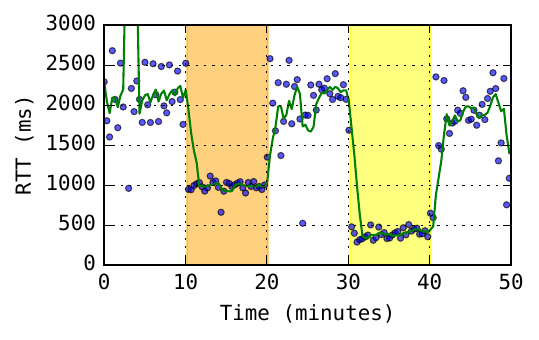}
        \caption{\emph{\Creepy}, 20\seconds interval, Wi-Fi}
    \end{subfigure}
    \hfil

    \vspace{0.2cm}

    \hfil
    \begin{subfigure}[b]{\figwidth\textwidth}
        \centering
        \includegraphics[width=\textwidth,trim={2.5mm 2.5mm 2.5mm 2.5mm},clip]{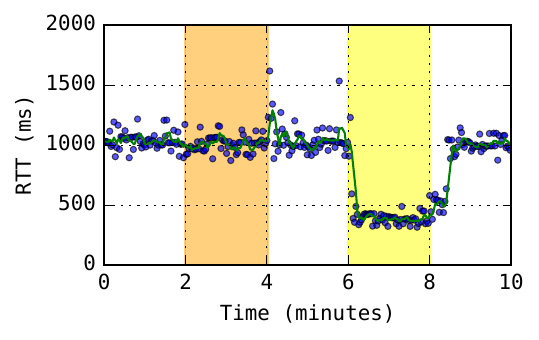}
        \caption{\emph{\Creepy}, 2\seconds interval, LTE}
    \end{subfigure}
    \hfil
    \begin{subfigure}[b]{\figwidth\textwidth}
        \centering
        \includegraphics[width=\textwidth,trim={2.5mm 2.5mm 2.5mm 2.5mm},clip]{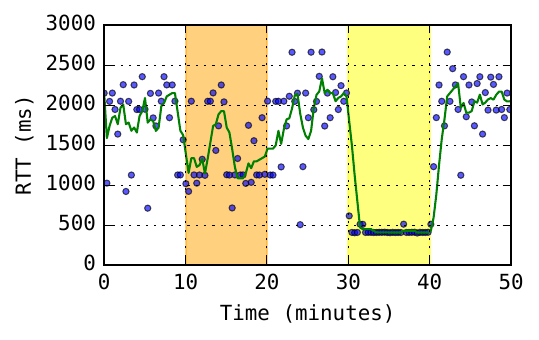}
        \caption{\emph{\Creepy}, 20\seconds interval, LTE}
    \end{subfigure}
    \hfil

    \vspace{0.5cm} %

    \hfil
    \begin{subfigure}[b]{\figwidth\textwidth}  
        \centering
        \includegraphics[width=\textwidth,trim={2.5mm 2.5mm 2.5mm 2.5mm},clip]{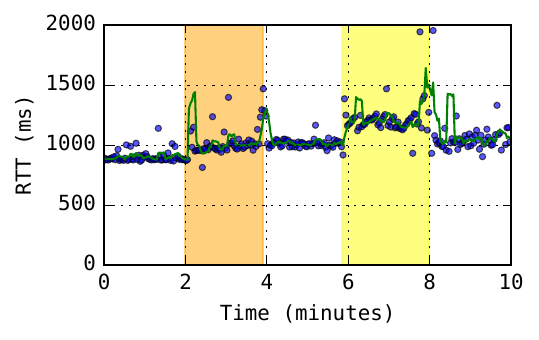}
        \caption{\emph{\Spooky}, 2\seconds interval, Wi-Fi}
        \label{fig:artefact1}
    \end{subfigure}
    \hfil
    \begin{subfigure}[b]{\figwidth\textwidth}
        \centering
        \includegraphics[width=\textwidth,trim={2.5mm 2.5mm 2.5mm 2.5mm},clip]{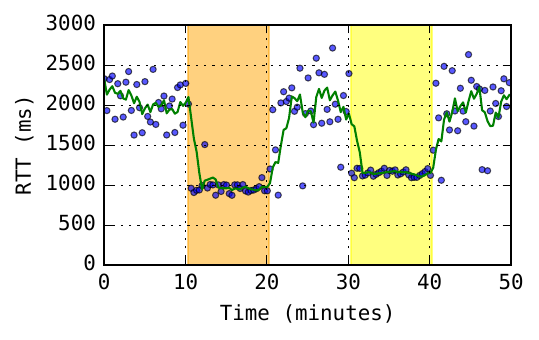}
        \caption{\emph{\Spooky}, 20\seconds interval, Wi-Fi}
    \end{subfigure}
    \hfil
    
    \vspace{0.2cm}

    \hfil
    \begin{subfigure}[b]{\figwidth\textwidth}
        \centering
        \includegraphics[width=1.05\textwidth,trim={2.5mm 2.5mm 2.5mm 2.5mm},clip]{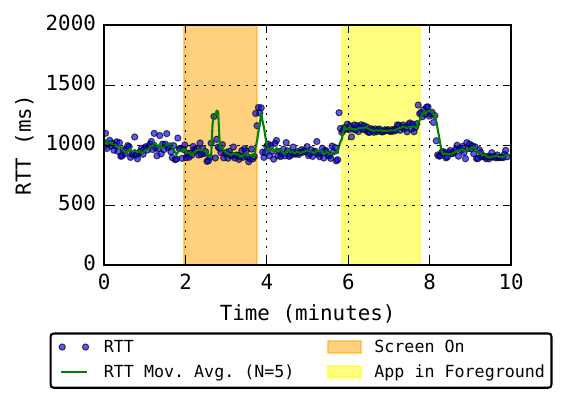}
        \caption{\emph{\Spooky}, 2\seconds interval, LTE}
    \end{subfigure}
    \hfil
    \begin{subfigure}[b]{\figwidth\textwidth}
        \centering
        \includegraphics[width=1.05\textwidth,trim={2.5mm 2.5mm 2.5mm 2.5mm},clip]{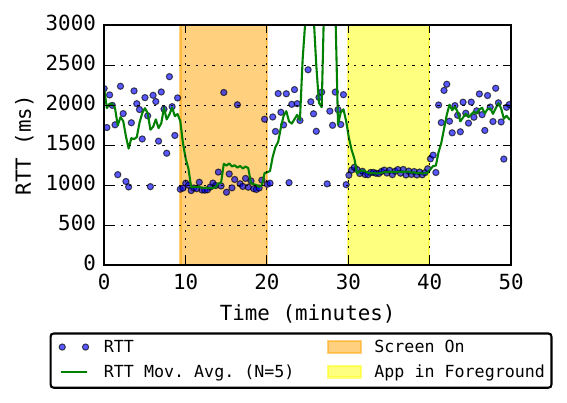}
        \caption{\emph{\Spooky}, 20\seconds interval, LTE}
        \label{fig:artefact2}
    \end{subfigure}
    \hfil
    \caption{Comparison of different probing intervals (2\seconds, 20\seconds), scenarios (\emph{\creepy}, \emph{\spooky}), and access technologies (Wi-Fi, LTE) for WhatsApp (measured on an iPhone 11)}
    \label{fig:iPhone-whatsapp-measurements}
\end{figure*}

\begin{figure*}[h]
    \centering    
    \hfil
    \begin{subfigure}[b]{\figwidth\textwidth}
        \centering
        \includegraphics[width=\textwidth,trim={2.5mm 2.5mm 2.5mm 2.5mm},clip]{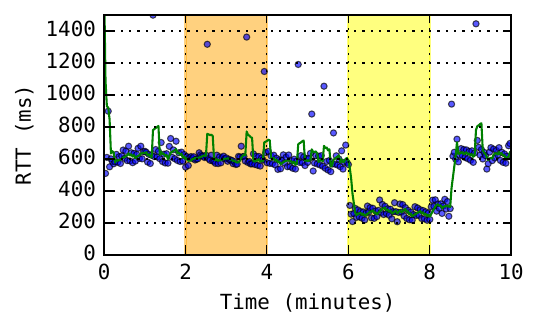}
        \caption{\emph{\Creepy}, 2\seconds interval, Wi-Fi}
    \end{subfigure}
    \hfil
    \begin{subfigure}[b]{\figwidth\textwidth}
        \centering
        \includegraphics[width=\textwidth,trim={2.5mm 2.5mm 2.5mm 2.5mm},clip]{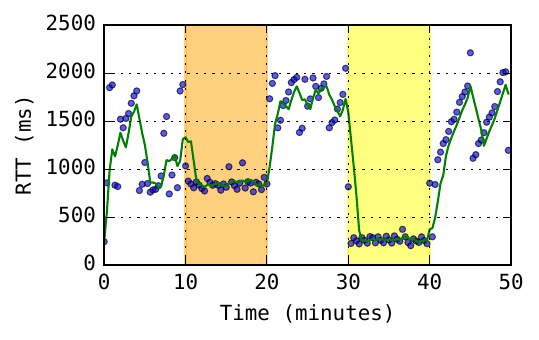}
        \caption{\emph{\Creepy}, 20\seconds interval, Wi-Fi}
    \end{subfigure}
    \hfil

    \vspace{0.2cm}

    \hfil
    \begin{subfigure}[b]{\figwidth\textwidth}
        \centering
        \includegraphics[width=\textwidth,trim={2.5mm 2.5mm 2.5mm 2.5mm},clip]{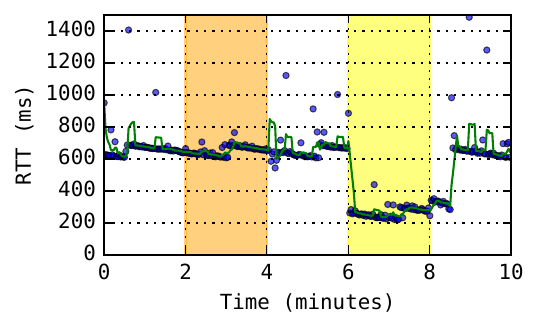}
        \caption{\emph{\Creepy}, 2\seconds interval, LTE}
    \end{subfigure}
    \hfil
    \begin{subfigure}[b]{\figwidth\textwidth}
        \centering
        \includegraphics[width=\textwidth,trim={2.5mm 2.5mm 2.5mm 2.5mm},clip]{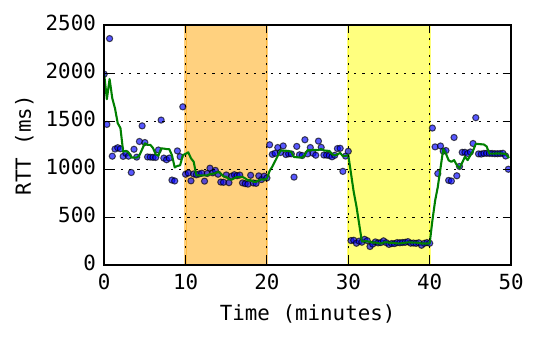}
        \caption{\emph{\Creepy}, 20\seconds interval, LTE}
    \end{subfigure}
    \hfil
    
    \vspace{0.5cm} %

    \hfil
    \begin{subfigure}[b]{\figwidth\textwidth}
        \centering
        \includegraphics[width=\textwidth,trim={2.5mm 2.5mm 2.5mm 2.5mm},clip]{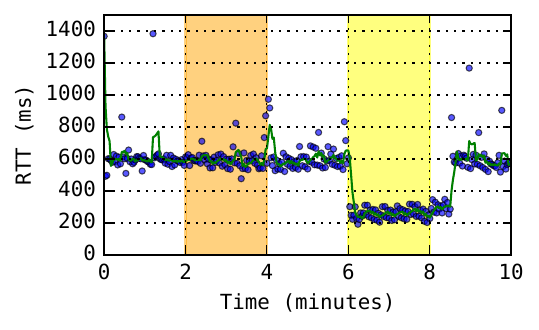}
        \caption{\emph{\Spooky}, 2\seconds interval, Wi-Fi}
    \end{subfigure}
    \hfil
    \begin{subfigure}[b]{\figwidth\textwidth}
        \centering
        \includegraphics[width=\textwidth,trim={2.5mm 2.5mm 2.5mm 2.5mm},clip]{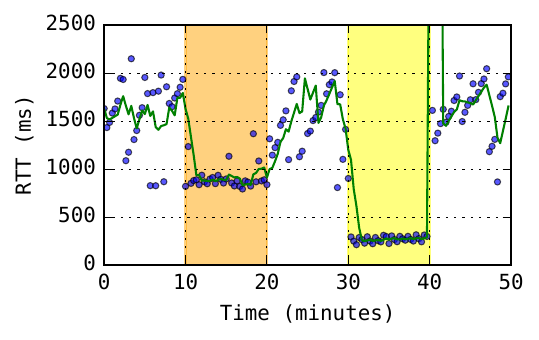}
        \caption{\emph{\Spooky}, 20\seconds interval, Wi-Fi}
    \end{subfigure}
    \hfil

    \vspace{0.2cm}

    \hfil
    \begin{subfigure}[b]{\figwidth\textwidth}
        \centering
        \includegraphics[width=1.05\textwidth,trim={2.5mm 2.5mm 2.5mm 2.5mm},clip]{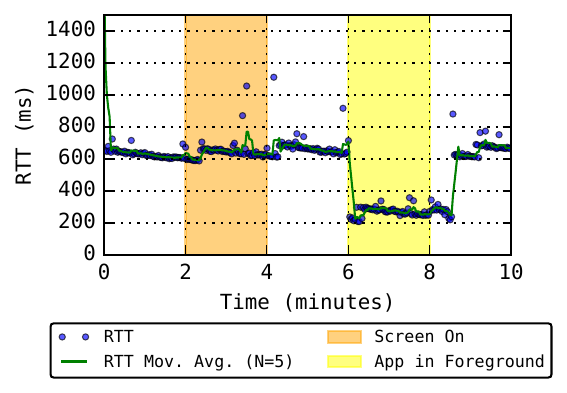}
        \caption{\emph{\Spooky}, 2\seconds interval, LTE}
    \end{subfigure}
    \hfil
    \begin{subfigure}[b]{\figwidth\textwidth}
        \centering
        \includegraphics[width=1.05\textwidth,trim={2.5mm 2.5mm 2.5mm 2.5mm},clip]{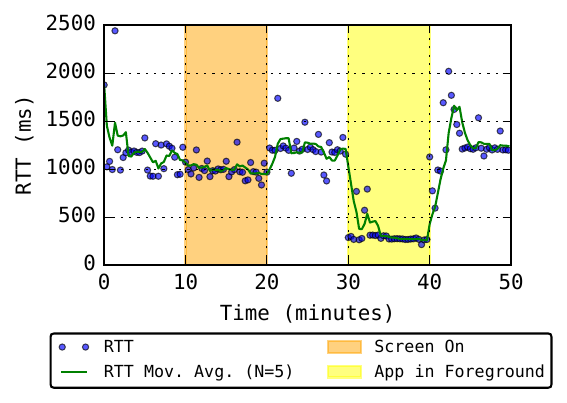}
        \caption{\emph{\Spooky}, 20\seconds interval, LTE}
    \end{subfigure}
    \hfil
    \caption{Comparison of different probing intervals (2\seconds, 20\seconds), scenarios (\emph{\creepy}, \emph{\spooky}), and access technologies (Wi-Fi, LTE) for Signal (measured on an iPhone 13 Pro)}
    \label{fig:iPhone-signal-measurements}
\end{figure*}

\end{document}